\documentclass[%
 reprint,
 amsmath,amssymb,
 aps
]{revtex4-2}

\usepackage{cancel}

\usepackage{soul}
\soulregister\ref{7}
\soulregister\cite{7}
\soulregister\eqref{7}
\soulregister\SI{7}
\soulregister\qty{7}
\usepackage{float}
\usepackage{amsmath}
\usepackage{upgreek}
\usepackage{exscale}
\usepackage{relsize}
\usepackage{graphicx}% Include figure files
\usepackage{dcolumn}% Align table columns on decimal point
\usepackage{bm}% bold math
\usepackage{mhchem}
\usepackage{ragged2e}
\usepackage{hyperref}
\hypersetup{
            hypertex=true,
            colorlinks=true,
            linkcolor=blue,
            anchorcolor=blue,
            citecolor=blue}

\usepackage{pdfpages}

\makeatletter
\AtBeginDocument{\let\LS@rot\@undefined}
\makeatother

\usepackage{pgffor}

\usepackage{siunitx}

\usepackage{color,soul}
\soulregister\cite7
 
\usepackage{xcolor}

\setstcolor{red}

\begin{document}

\preprint{APS/123-QED}

\title{Cascade of Modal Interactions in Nanomechanical Resonators with Soft Clamping}

\author{Zichao Li\textsuperscript{*,1},
        Minxing Xu\textsuperscript{1,2},
        Richard A. Norte\textsuperscript{1,2},
        Alejandro M. Aragón\textsuperscript{1},
        Peter G. Steeneken\textsuperscript{1,2}
        and Farbod Alijani\textsuperscript{*,1}
        }
 
\affiliation{%
 {\rm {\textsuperscript{1}}}Department of Precision and Microsystems Engineering, Delft University of Technology, Mekelweg 2, 2628 CD Delft, The Netherlands\\
 {\rm {\textsuperscript{2}}}Kavli Institute of Nanoscience, Delft University of Technology, Lorentzweg 1, 2628 CJ Delft, The Netherlands
}%

\begin{abstract}
We uncover a chain of nonlinear modal interactions in softly clamped nanostring resonators. The process involves the sequential coupling of five mechanical modes, during frequency sweeps, yielding a broad nonlinear response with nearly constant amplitude. We demonstrate that soft clamping enables this cascaded energy transfer and amplifies the effective geometric nonlinearity of the driven mode by an order of magnitude. Analytical and finite element-based reduced-order models capture the key features of the coupling cascade and clarify its underlying mechanism. The phenomenon is generic in nonlinear vibrational systems and can be tailored through soft-clamping design strategies.

\end{abstract}

\maketitle

%\tableofcontents

\newpage

%\section{I. Introduction}

Complex behaviors across physical systems, from fluid flows to biological synchrony, often arise when a change in a system parameter triggers a cascade of interconnected phenomena~\cite{strogatz2012sync,gu2025emergence,japaridze2025synchronization}. 
Such cascaded interactions are not peripheral in nonlinear dynamical systems; they are central to how energy and information propagate across coupled degrees of freedom, giving rise to rich dynamical patterns and abrupt transitions~\cite{lorenz2000butterfly,may1976simple,vakakis2022nonlinear,chen2017direct}. 

In recent years, micro- and nanomechanical systems have served as ideal experimental platforms for exploring nonlinear dynamics, owing to their high susceptibility to large-amplitude oscillations. These systems provide access to regimes where exotic dynamical states can emerge~\cite{matheny2019exotic,guttinger2017energy,Kecskekler2021tuning,yang2021persistent,houri2020generic,eriksson2023controllable,belardinelli2025hidden}.
A key parameter along this pathway is the coupling between vibrational modes, which can link distinct motion states and open up opportunities for frequency stabilization~\cite{antonio2012frequency,shoshani2024extraordinary}, energy harvesting~\cite{chen2023internal,asadi2018nonlinear}, and frequency comb generation~\cite{keskekler2022symmetry,sun2023generation,fu2025sideband,wu2025self}. Mode coupling in a resonant mechanical system occurs when there is a substantial energy transfer rate between different vibrational modes~\cite{yang2015experimental}.
However, to date, studies of mode coupling have predominantly focused on interactions between two modes, typically under conditions known as internal resonance ~\cite{keskekler2022symmetry,antonio2012frequency,qiao2023frequency,fan2024internal,monteil2014nonlinear}. Studies of simultaneous mode coupling involving more than two modes have been rare because realizing multiple modal interactions generally requires both commensurate frequency ratios and sufficient energy to simultaneously activate additional modes alongside the driven one.  

Here we overcome these limitations and show that cascaded modal interactions can naturally arise and be harnessed in nanoresonators with soft clamping~\cite{li2023tuning, Spiderweb, fedorov2020fractal,li2024strain}. Our experiments reveal that successive intermodal couplings enable quasi-constant oscillation amplitudes over a broad frequency range. To elucidate this effect, we use analytical models and finite element (FE)-based reduced-order model (ROM) simulations to quantify the coupling strengths among different mechanical modes. We demonstrate that these cascaded interactions amplify the effective Duffing constant of the resonator by more than an order of magnitude, effectively flattening the amplitude–frequency response of the driven mode. This establishes cascaded modal coupling as a powerful mechanism for both amplitude stabilization and programmable multistable responses, offering new opportunities for precision control of signal fidelity and frequency stability in nanomechanical resonators.

\begin{figure*}
\includegraphics[scale=0.9]{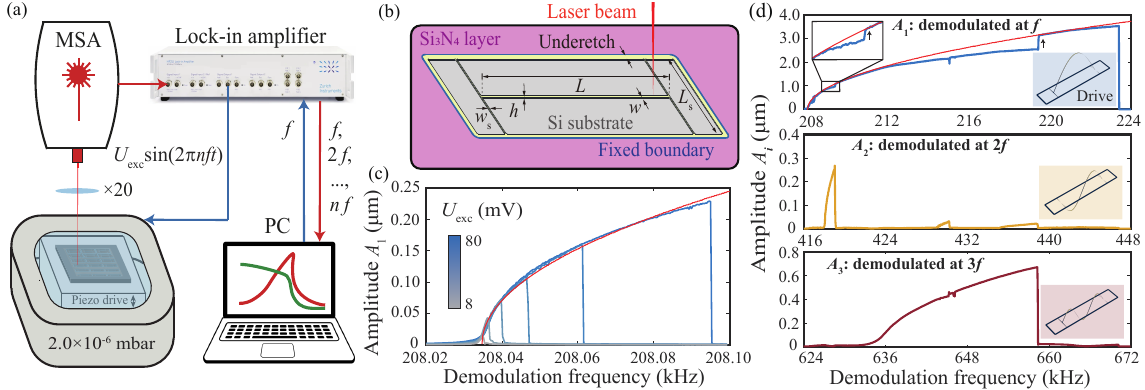}
\caption{\label{fig.1} \textbf{Measurement of mode coupling in a nanomechanical string resonator with soft-clamping supports.} (a) Schematic of the measurement set-up comprising an MSA400 laser Doppler vibrometer (LDV) for reading out the motion at different harmonics of the drive frequency ($f$, 2$f$, ..., $n$$f$) and a piezo-actuator for generating the excitation. (b) The geometric design parameters of a \ce{Si3N4} nanostring resonator with soft-clamping supports. (c) Duffing nonlinear response curves of the first mode of the device with $L_{\rm s} = \qty[]{50}{\micro\metre}$, under different drive levels without mode coupling. (d) Nonlinear response curves of the same device under a stronger drive level ($U_{\rm exc} = \qty[]{6}{\volt}$). The second (yellow) and third (ochre) modes are both activated by mode coupling.  The red lines in (c) and (d) (first panel) are the backbone curves of the first mode.}
\end{figure*}

Our measurements are performed on \ce{Si3N4} nanostrings (thickness $h = \SI{90}{\nano\metre}$, pre-stress $\sigma_0=\qty[]{1.06}{\giga\pascal}$) featuring slender support beams at the boundaries to mediate soft clamping (see Supplemental Material S1~\cite{supplemental}~\nocite{xu2024high,villanueva2014evidence,hauer2013general,li2024strain,miller2018effective,nayfeh2008nonlinear,li2023tuning}). Such designs lead to high $Q$ factors as well as approximate integer ratio eigenfrequencies ~\cite{li2024strain,li2025finite} and thus can promote efficient intermodal couplings. Fig.~\ref{fig.1}a shows our measurement setup, with Fig.~\ref{fig.1}b illustrating the geometric parameters of one fabricated device. All devices have identical dimensions ($L = \qty[]{200}{\micro\metre}$, $w = \qty[]{2}{\micro\metre}$, and $w_{\rm s} = \qty[]{1}{\micro\metre}$), but differ in support length $L_{\rm s}$. We note that the presence of soft-clamping supports tunes the in-plane stress in the central string~\cite{li2023tuning,li2024strain}.

To characterize the nonlinear dynamics of these nanostrings, we fix the chip patterned with suspended resonators to a piezo actuator that provides a harmonic base excitation in the out-of-plane direction. We use a Zurich Instruments HF2LI lock-in amplifier to perform frequency sweeps in the spectral neighborhood of the first resonance, together with a Polytec MSA400 laser Doppler vibrometer (LDV) for detecting the out-of-plane vibrations of our devices. All modes discussed in this work refer specifically to out-of-plane modes. The measurement laser is focused at a position $L$/12 from the support on the central string, ensuring it is distant from nodal points of the lowest vibrational modes (see Supplemental Material S2~\cite{supplemental}). We perform all measurements at room temperature in a vacuum chamber with a pressure below $\SI{2e-6}{\milli\bar}$ to minimize air damping. 

To probe the geometric nonlinearity of our devices, we perform forward frequency sweeps at different drive levels and measure the vibrations of the central string~\cite{li2024strain}. As an example, Fig.~\ref{fig.1}c shows the frequency responses at various drive voltages, ranging from $U_{\rm exc} = \qty[]{8}{\milli\volt}$ to $\qty[]{80}{\milli\volt}$, for a device with $L_{\rm s} = \qty[]{50}{\micro\metre}$. To quantify the strength of nonlinearity, we use the Duffing equation:
\begin{equation}
\begin{aligned}
\ddot q_1 + c_1 \dot q_1 + \omega_1^2 q_1 + \beta_1 q_1^3 = F_{\rm exc} \sin{(2 \pi f t)},
\label{equation.1}
\end{aligned}
\end{equation}
\addtocounter{equation}{0}

\noindent where $q_1$ is the generalized coordinate of the first mode, $F_{\rm exc} \sin{(2 \pi f t)}$ is its effective harmonic drive with excitation frequency $f$ from the piezo. Furthermore, $\omega_1 = 2 \pi f_1$, $c_1 = \omega_1 / Q_1$ and $\beta_1$ are the angular resonance frequency, mass-normalized damping coefficient, and Duffing constant, respectively, where $Q_1=6.07 \times 10^5$ is the quality factor characterized by the ring-down measurement~\cite{li2023tuning}. We extract $\beta_1$ by fitting the backbone curve (red line in Fig.~{\ref{fig.1}}c) ~\cite{nayfeh2008nonlinear,schmid2016fundamentals}. 

In Fig.~\ref{fig.1}d, we show the frequency response of the same device measured in Fig.~\ref{fig.1}c driven at a stronger excitation level ($U_{\rm exc} = \qty[]{6}{\volt}$) around the first resonance. Apart from the signal demodulated with the drive frequency $f$ (blue), we also detect higher harmonics demodulated at 2$f$ (yellow) and 3$f$ (ochre). We note that the frequency response of the first mode deviates from the backbone curve of the Duffing response (red line in Fig.~\ref{fig.1}d) when the higher harmonics are detected. Since the resonance frequencies of higher modes of a string resonator are close to integer multiples of the first, we attribute the deviation from the backbone curve to modal interactions between the first and higher-order modes~\cite{houri2020demonstration}. When the oscillations of the coupled modes drop, the energy stored in the higher modes transfers back to the driven mode, bringing its amplitude closer to its backbone (see arrows in Fig.~\ref{fig.1}d). We provide a more detailed explanation of the multiple disconnected features that appear in the response near $2f$ in Supplemental Material S3~\cite{supplemental}.

\begin{figure}[t]
\centering
\includegraphics[scale=0.82]{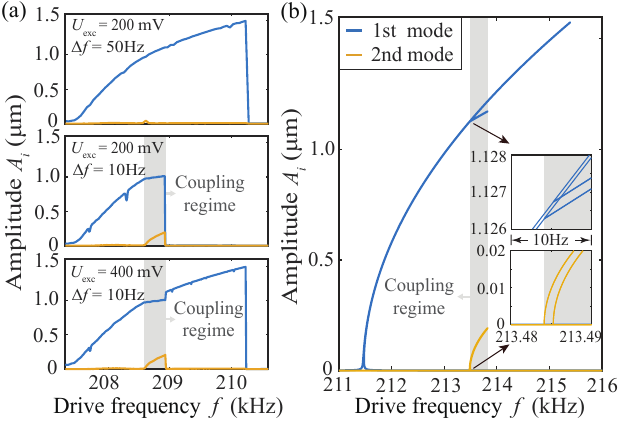}
\caption{\label{fig.2} \textbf{Different response branches of the first mode determined by its coupling to the second mode.} (a) Measured response curves by driving the first mode under different $U_{\rm exc}$ and $\Delta f$, showing frequency demodulation around $f$ (blue, first mode) and $2f$ (yellow, second mode). (b) Simulated response curves based on an FE-based ROM, showing the directly driven first mode near $f$ (blue) and the coupling-induced second mode around $2f$ (yellow). The gray area marks the frequency range where the second mode is activated.}
\end{figure}

\begin{figure*}
\includegraphics[scale=0.92]{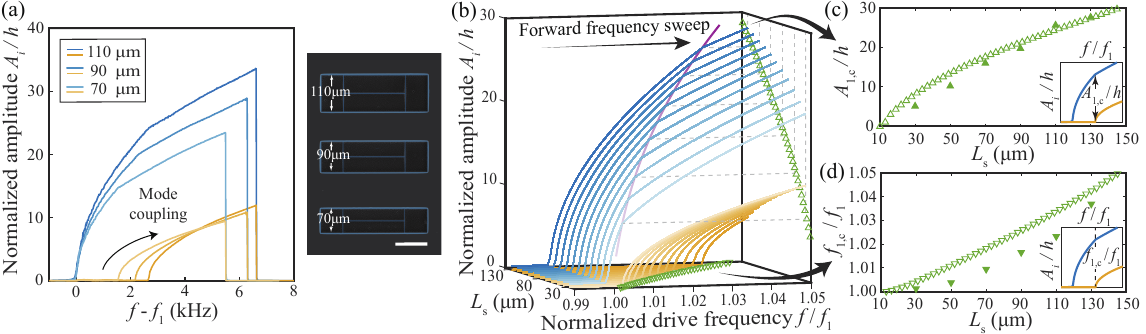}
\caption{\label{fig.3} \textbf{Influence of the soft-clamping supports on the coupled dynamics of the lowest two modes.} (a) Measured response curves of the lowest two modes of string resonators with three different $L_{\rm s}$, showing frequency demodulation around $f$ (blue, first mode) and $2f$ (yellow, second mode). The colors of the curves gradually fade as $L_{\rm s}$ decreases. The SEM image shows the measured devices (colored in blue). The white bar is $\qty[]{100}{\micro\metre}$. (b) Simulated response curves using the FE-based ROMs of devices with varying $L_{\rm s}$, showing the directly driven first mode near $f$ (blue) and the coupling-induced second mode around $2f$ (yellow). The upward and downward hollow triangles represent the onset frequency of the coupled mode ($f_{\rm 1,c}/f_1$) and the corresponding amplitude ($A_{\rm 1,c}/h$), respectively, as predicted by Eq.~\eqref{equation.5}. The purple line traces the onsets of modal interactions, corresponding to the kinks in the blue curves. (c)(d) Comparison of the simulated (hollow triangles) and measured (solid triangles) onset amplitude $A_{\rm 1,c}/h$ and frequency $f_{\rm 1,c}/f_1$. The insets define $f_{\rm 1,c}/f_1$ and $A_{\rm 1,c}/h$, respectively. }
\end{figure*}

To further examine the amplitude jumps (arrows) and coupling conditions in Fig.~\ref{fig.1}d, we perform a detailed investigation on the coupling between the first and second modes in a device with identical geometry, as shown in Fig.~\ref{fig.2}a. Starting with a drive voltage of $U_{\rm exc} = \qty[]{200}{\milli\volt}$ and a frequency step of $\Delta f = \qty[]{50}{\hertz}$, we obtain a Duffing response of the first mode without activation of the second mode. However, when we reduce $\Delta f$ to $\qty[]{10}{\hertz}$ while keeping the same drive voltage, we observe that the second mode is activated (gray area). As the drive frequency is swept forward, the oscillations of both modes drop to their respective lower amplitude branches. To further explore the influence of drive level, we double $U_{\rm exc}$ to $\qty[]{400}{\milli\volt}$. We can see the oscillation of the first mode is driven back to its upper stable branch at the end of the coupling regime, while the second mode drops to its lower branch, consistent with Fig.~\ref{fig.1}d. This observation suggests that the driving parameters can influence the slope of the solution branches during a frequency sweep.

To verify our observation, and to study the mode coupling mechanism in a two-mode system, we employ an FE model of the nanoresonator and build a two-degree-of-freedom (2-DOF) ROM that comprises the first two modes (see previous works~\cite{muravyov2003determination,li2024strain,kecskekler2023multimode} and Supplemental Material S3~\cite{supplemental} for more details):
\begin{subequations}
\begin{align}
\ddot{q_1} + c_1 \dot{q_1} + \omega_1^2 q_1 + \beta_1 q_1^3 + \gamma q_1 q_2^2 &= F_{\rm exc} \sin{(2 \pi f t)}, \label{equation.2a}\\
\ddot{q_2} + c_2 \dot{q_2} + \omega_2^2 q_2 + \beta_2 q_2^3 + \gamma q_1^2 q_2 &= F_{\rm exc}^{\prime} \sin{(4 \pi f t)}.
\label{equation.2b}
\end{align}
\end{subequations}
\addtocounter{equation}{0}

\noindent Here, $q_2$ is the generalized coordinate of the second mode. $\omega_2 = 2 \pi f_2$, $c_2 = \omega_2 / Q_2$ and $\beta_2$ are its angular resonance frequency, mass-normalized damping coefficient and Duffing constant, respectively, where $Q_2$ is the quality factor of the second mode. 
We introduce a minute harmonic force $F_{\rm exc}^{\prime} \sin(4\pi f t)$ to capture the influence of the quadratic coupling induced by symmetry-breaking imperfections of the device. This term is necessary to excite the second mode (see Supplementary Material S3). Moreover, $\gamma$ in Eqs.~\eqref{equation.2a} and~\eqref{equation.2b} represents the mass-normalized dispersive coupling coefficient between the two modes that is directly obtained from the ROM and promotes energy transfer~\cite{yang2015experimental}. We use numerical continuation~\cite{Matcont} to compute all possible solution branches of the 2-DOF system, as shown in Fig.~\ref{fig.2}b (see Supplemental Material S6~\cite{supplemental}). At the onset of mode coupling we also notice the emergence of a second solution branch with a slope different from the common Duffing response in the coupling regime~\cite{hellbach2024nonlinearity}, which matches our experimental observation obtained by frequency sweeps. The simulation reveals that the amplitude of the second mode increases rapidly at the onset of mode coupling (see insets in Fig.~\ref{fig.2}b), implying a coarse frequency sweep may overlook its activation window. We attribute this phenomenon to the high $Q$-factor of our devices and the presence of nonlinear coupling terms ($\gamma q_1 q_2^2$ and $\gamma q_1^2 q_2$), which govern the energy exchange between the modes in this 2-DOF system. In subsequent experiments, we ensure sufficiently small frequency steps to reliably activate the coupled motion and trace the solution branch that emerges from the coupling. 

Next, to quantify the influence of soft-clamping supports on mode coupling, we measure devices with varying support lengths $L_{\rm s}$ undergoing intermodal coupling between the lowest two modes, as shown in Fig.~\ref{fig.3}a (see Supplemental Material S5~\cite{supplemental} for more measurement results). The kinks at relatively large amplitudes signify transitions to alternative solution branches---features that closely resemble those observed in our numerical simulations in Fig.~\ref{fig.2}b, and are indicative of extra vibrational modes becoming active through dispersive coupling. 

To derive the onset conditions of a two-mode coupling, we apply Harmonic Balance Method (HBM) (see Supplemental Material S3~\cite{supplemental}) to Eqs.~\eqref{equation.2a} and~\eqref{equation.2b}, thereby obtaining analytical expressions that predict the frequency $f_{\rm 1,c}$ and amplitude $A_{\rm 1,c}$ of the first mode at the onset of coupling (see kinks in Fig.~\ref{fig.3}a):
\begin{equation}
\begin{aligned}
f_{\rm 1,c} = \frac{1}{2 \pi} \sqrt{\frac{\omega_2^2 - \frac{2 \gamma}{3 \beta_1} \omega_1^2 }{4 - \frac{2 \gamma}{3 \beta_1}}}, A_{\rm 1,c} = \sqrt{\frac{\omega_2^2-4 \omega_1^2}{3 \beta_1 - \frac{1}{2} \gamma}}.
\label{equation.5}
\end{aligned}
\end{equation}
\addtocounter{equation}{0}

\noindent We use Eq.~\eqref{equation.5} and parameters
from FE-based 2-DOF ROMs to predict the onset of coupling (see green triangles in Fig.~\ref{fig.3}b). The obtained values closely match the kinks observed in response curves simulated from the same ROMs (see Supplemental Material S6~\cite{supplemental}). The predictions also match $f_{\rm 1,c}$ and $A_{\rm 1,c}$ extracted from experimental response curves for devices with different support lengths, as demonstrated in Figs.~\ref{fig.3}c and d. These results confirm the validity of Eq.~\eqref{equation.5} in predicting the kink in the nonlinear frequency response curve.

Apart from the onset of two-mode coupling, we can see from Fig.~\ref{fig.3}a that the coupled response also has an impact on the spring-hardening nonlinearity of the driven mode. The amplitude-frequency relationship after the kinks remains parabolic, which closely tracks the effective backbone curve of the first mode undergoing dispersive coupling with the second mode (see Supplemental Material S3~\cite{supplemental}).

The HBM analytical framework further allows us to extend the two-mode analysis to a situation where multiple modes are coupled via cubic nonlinearity and excited simultaneously, in a cascade-like fashion, when only the lowest mode is driven. To that end, and in order to capture the impact of successive dispersive couplings from the interaction potential $\propto  q_1^2q^2_i$ ($i \geq 2$) on the nonlinear dynamics of the fundamental mode, we derive a recursive relation for the effective Duffing constant as additional coupled modes are introduced (see Supplemental Material S7~\cite{supplemental}):
\begin{equation}
\begin{aligned}
\beta_{\rm 1,eff}^{(i)} = \frac{3}{4} \beta_{\rm 1,eff}^{(i-1)} + \frac{3 i^2 \beta_{\rm 1,eff}^{(i-1)} \gamma_{1,i} - 2 \gamma_{1,i}^2}{6 \beta_i - 4 i^2 \gamma_{1,i}},\label{equation.7}
\end{aligned}
\end{equation}
\addtocounter{equation}{0}
\noindent where $\beta_{\rm 1,eff}^{(i)}$ is the effective Duffing constant by including up to the $i$th mode ($i \geq 2$, $\beta_{\rm 1,eff}^{(1)} = \beta_1$), $\beta_i$ is the intrinsic Duffing constant of the $i$th mode, and $\gamma_{1,i}$ is the dispersive coupling coefficient between the first and $i$th modes. This relation serves as a predictive map that quantifies how successive dispersive interactions modulate the nonlinearity of the first mode, providing both physical insight and a tool for engineering multi-mode dispersive interactions. However, we note that this approximation neglects the interactions among the coupled modes other than the first mode, and assumes that higher-order modes are activated (See Supplemental Material S7~\cite{supplemental}).

In Fig.~\ref{fig.4}a, we schematically illustrate this effect: the red curve shows the reshaped backbone of the first mode’s response as successive dispersive couplings are introduced. The initial segment represents the uncoupled response of the first mode. Kinks in the curve, marked by black dots, indicate changes in the slope of the Duffing response and correspond to higher-order modes coupling into the dynamics. In fact, these sharp transitions reflect how higher modes act as energy reservoirs, absorbing part of the energy injected into the first mode and thereby suppressing its vibration amplitude~\cite{antonio2012frequency,chen2017direct}.

\begin{figure}[t]
\centering
\includegraphics[scale=0.9]{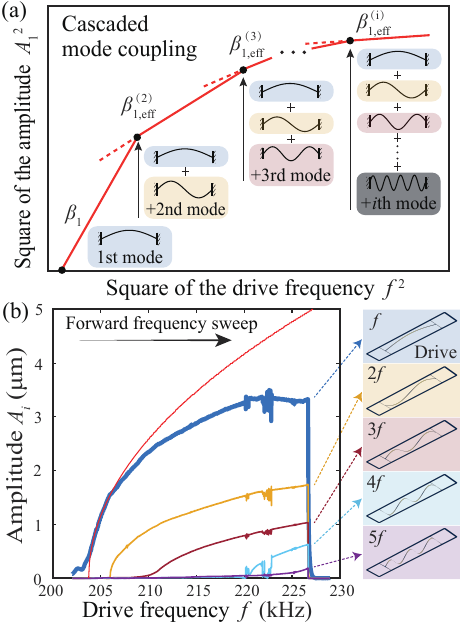}
\caption{\label{fig.4} \textbf{Cascaded interactions in a nanostring.} (a) The red line denotes the driven mode’s effective backbone, with higher modes triggered at its kinks via dispersive coupling. (b) Measured response curves under five-mode couplings in the device with $L_{\rm s} = \qty[]{50}{\micro\metre}$. The bold blue line is the frequency response demodulated by the drive frequency $f$, while the others represent the signals demodulated by 2$f$ (yellow), 3$f$ (ochre), 4$f$ (cyan) and 5$f$ (purple), respectively. The mode shapes from FE analysis are shown on the right. The red line is the fitted backbone curve of the first mode before mode couplings initiate.}
\end{figure}

By substituting parameters extracted from FE-based ROMs for the device measured in Fig.~\ref{fig.1}, we find that incorporating two-mode coupling leads to a 52$\%$ increase in the effective Duffing constant. With interactions among five modes, the model of Eq.~\eqref{equation.7} predicts an enhancement exceeding a 26-fold increase in $\beta_{\rm 1,eff}^{(i)}$ (see Supplemental Material S7~\cite{supplemental}).

To experimentally investigate the cascaded interactions predicted by our analytical model (Eq.~\eqref{equation.7}), we perform forward frequency sweeps on the device measured in Fig.~\ref{fig.1} with a higher excitation level (\(U_{\rm exc} = \qty[]{20}{\volt}\)) to activate higher-order modes. 
Consistent with the analytical prediction, the measured frequency response in Fig.~\ref{fig.4}b shows a clear suppression of the fundamental mode’s amplitude, resulting in an almost constant response over a broad frequency range as higher-order modes participate in the coupled dynamics. The unstable fluctuations observed after the onset of the fourth mode are attributed to the activation of coupling to higher-order modes ($i>6$). To assess the validity of Eq.~\eqref{equation.7} in capturing the rescaling of the Duffing constant via modal coupling, we also performed FE-based ROM simulations including all-to-all coupling terms (see Supplemental Material S7~\cite{supplemental}). These simulations reveal that while higher-order resonant interactions are present and affect the overall dynamics, the dominant dispersive terms influencing the first mode are those already included in the analytical model.

It is worth noting that, despite the significant potential of intermodal modulation, achieving multimodal coupling in most micro- and nanomechanical resonators is inherently challenging due to the incommensurate ratios of their eigenfrequencies. Soft clamping offers a key advantage in this regard: its inherent design flexibility enables precise tuning of eigenfrequency ratios. Under hardening nonlinearity, this allows rapid access to the coupling regime during a forward frequency sweep, without the need for large vibrational amplitudes to overcome detuning. Moreover, soft clamping significantly enhances $Q$ factor, which promotes the participation of a larger number of vibrational modes by lowering the energy threshold required to initiate modal interactions.

To summarize, we present evidence and models for cascaded intermodal couplings in resonators with soft clamping. The ability of using these cascades for exciting multiple couplings in high-$Q$ resonators with a single drive tone not only facilitates the actuation of successive modes, but also allows reshaping and engineering the system's nonlinear dynamic response. 
Cascades of intermodal couplings can stabilize the amplitude of the driven mode across a broad frequency range (see Fig.~{\ref{fig.4}}b), and allow more freedom in tuning the amplitude-frequency response than when using only two-mode couplings. This multi-mode nonlinear response engineering might, with more refined procedures that still need to be developed, enable amplitude and frequency stabilization, as well as potentially achieve phase noise reduction in nonlinear oscillators beyond what has been demonstrated for two-mode couplings~\cite{antonio2012frequency,defoort2022amplitude,villanueva2011nanoscale}. 
More advanced prospects of multi-mode cascades involve their use for nanomechanical error correction in a single resonator, along a route that has recently been demonstrated using three separate coupled oscillators~\cite{jin2025nanomechanical}. Finally, programmable multistability of all, or a subset of, cascade modes by appropriate drive control could open new opportunities for logic operations and nanomechanical computation~\cite{mahboob2008bit,bagheri2011dynamic,yao2014logic}. \\[1pt]

\section{Acknowledgements}
Funded/Co-funded by the European Union (ERC Consolidator, NCANTO, 101125458). Z.L. also acknowledges financial support from China Scholarship Council and the fruitful discussions with Dr. Ata Ke{\c{s}}kekler. This work is also part of the project ``Probing the physics of exotic superconductors with microchip Casimir experiments (740.018.020)'' of the research program NWO Start-up, which is partly financed by the Dutch Research Council (NWO). M.X. and R.A.N. acknowledge support from the Kavli Nanolab Delft.

\bibliography{apssamp}% Produces the bibliography via BibTeX.

\providecommand{\noopsort}[1]{}\providecommand{\singleletter}[1]{#1}%
\begin{thebibliography}{49}%
\makeatletter
\providecommand \@ifxundefined [1]{%
 \@ifx{#1\undefined}
}%
\providecommand \@ifnum [1]{%
 \ifnum #1\expandafter \@firstoftwo
 \else \expandafter \@secondoftwo
 \fi
}%
\providecommand \@ifx [1]{%
 \ifx #1\expandafter \@firstoftwo
 \else \expandafter \@secondoftwo
 \fi
}%
\providecommand \natexlab [1]{#1}%
\providecommand \enquote  [1]{``#1''}%
\providecommand \bibnamefont  [1]{#1}%
\providecommand \bibfnamefont [1]{#1}%
\providecommand \citenamefont [1]{#1}%
\providecommand \href@noop [0]{\@secondoftwo}%
\providecommand \href [0]{\begingroup \@sanitize@url \@href}%
\providecommand \@href[1]{\@@startlink{#1}\@@href}%
\providecommand \@@href[1]{\endgroup#1\@@endlink}%
\providecommand \@sanitize@url [0]{\catcode `\\12\catcode `\$12\catcode `\&12\catcode `\#12\catcode `\^12\catcode `\_12\catcode `\%12\relax}%
\providecommand \@@startlink[1]{}%
\providecommand \@@endlink[0]{}%
\providecommand \url  [0]{\begingroup\@sanitize@url \@url }%
\providecommand \@url [1]{\endgroup\@href {#1}{\urlprefix }}%
\providecommand \urlprefix  [0]{URL }%
\providecommand \Eprint [0]{\href }%
\providecommand \doibase [0]{https://doi.org/}%
\providecommand \selectlanguage [0]{\@gobble}%
\providecommand \bibinfo  [0]{\@secondoftwo}%
\providecommand \bibfield  [0]{\@secondoftwo}%
\providecommand \translation [1]{[#1]}%
\providecommand \BibitemOpen [0]{}%
\providecommand \bibitemStop [0]{}%
\providecommand \bibitemNoStop [0]{.\EOS\space}%
\providecommand \EOS [0]{\spacefactor3000\relax}%
\providecommand \BibitemShut  [1]{\csname bibitem#1\endcsname}%
\let\auto@bib@innerbib\@empty
%</preamble>
\bibitem [{\citenamefont {Strogatz}(2012)}]{strogatz2012sync}%
  \BibitemOpen
  \bibfield  {author} {\bibinfo {author} {\bibfnamefont {S.~H.}\ \bibnamefont {Strogatz}},\ }\href@noop {} {\emph {\bibinfo {title} {Sync: How order emerges from chaos in the universe, nature, and daily life}}}\ (\bibinfo  {publisher} {Hachette UK},\ \bibinfo {year} {2012})\BibitemShut {NoStop}%
\bibitem [{\citenamefont {Gu}\ \emph {et~al.}(2025)\citenamefont {Gu}, \citenamefont {Guiselin}, \citenamefont {Bain}, \citenamefont {Zuriguel},\ and\ \citenamefont {Bartolo}}]{gu2025emergence}%
  \BibitemOpen
  \bibfield  {author} {\bibinfo {author} {\bibfnamefont {F.}~\bibnamefont {Gu}}, \bibinfo {author} {\bibfnamefont {B.}~\bibnamefont {Guiselin}}, \bibinfo {author} {\bibfnamefont {N.}~\bibnamefont {Bain}}, \bibinfo {author} {\bibfnamefont {I.}~\bibnamefont {Zuriguel}},\ and\ \bibinfo {author} {\bibfnamefont {D.}~\bibnamefont {Bartolo}},\ }\href@noop {} {\bibfield  {journal} {\bibinfo  {journal} {Nature}\ }\textbf {\bibinfo {volume} {638}},\ \bibinfo {pages} {112} (\bibinfo {year} {2025})}\BibitemShut {NoStop}%
\bibitem [{\citenamefont {Japaridze}\ \emph {et~al.}(2025)\citenamefont {Japaridze}, \citenamefont {Struijk}, \citenamefont {Swamy}, \citenamefont {Ros{\l}o{\'n}}, \citenamefont {Shoshani}, \citenamefont {Dekker},\ and\ \citenamefont {Alijani}}]{japaridze2025synchronization}%
  \BibitemOpen
  \bibfield  {author} {\bibinfo {author} {\bibfnamefont {A.}~\bibnamefont {Japaridze}}, \bibinfo {author} {\bibfnamefont {V.}~\bibnamefont {Struijk}}, \bibinfo {author} {\bibfnamefont {K.}~\bibnamefont {Swamy}}, \bibinfo {author} {\bibfnamefont {I.}~\bibnamefont {Ros{\l}o{\'n}}}, \bibinfo {author} {\bibfnamefont {O.}~\bibnamefont {Shoshani}}, \bibinfo {author} {\bibfnamefont {C.}~\bibnamefont {Dekker}},\ and\ \bibinfo {author} {\bibfnamefont {F.}~\bibnamefont {Alijani}},\ }\href@noop {} {\bibfield  {journal} {\bibinfo  {journal} {Small}\ }\textbf {\bibinfo {volume} {21}},\ \bibinfo {pages} {2407832} (\bibinfo {year} {2025})}\BibitemShut {NoStop}%
\bibitem [{\citenamefont {Lorenz}(2000)}]{lorenz2000butterfly}%
  \BibitemOpen
  \bibfield  {author} {\bibinfo {author} {\bibfnamefont {E.}~\bibnamefont {Lorenz}},\ }\href@noop {} {\bibfield  {journal} {\bibinfo  {journal} {World Scientific Series on Nonlinear Science Series A}\ }\textbf {\bibinfo {volume} {39}},\ \bibinfo {pages} {91} (\bibinfo {year} {2000})}\BibitemShut {NoStop}%
\bibitem [{\citenamefont {May}(1976)}]{may1976simple}%
  \BibitemOpen
  \bibfield  {author} {\bibinfo {author} {\bibfnamefont {R.~M.}\ \bibnamefont {May}},\ }\href@noop {} {\bibfield  {journal} {\bibinfo  {journal} {Nature}\ }\textbf {\bibinfo {volume} {261}},\ \bibinfo {pages} {459} (\bibinfo {year} {1976})}\BibitemShut {NoStop}%
\bibitem [{\citenamefont {Vakakis}\ \emph {et~al.}(2022)\citenamefont {Vakakis}, \citenamefont {Gendelman}, \citenamefont {Bergman}, \citenamefont {Mojahed},\ and\ \citenamefont {Gzal}}]{vakakis2022nonlinear}%
  \BibitemOpen
  \bibfield  {author} {\bibinfo {author} {\bibfnamefont {A.~F.}\ \bibnamefont {Vakakis}}, \bibinfo {author} {\bibfnamefont {O.~V.}\ \bibnamefont {Gendelman}}, \bibinfo {author} {\bibfnamefont {L.~A.}\ \bibnamefont {Bergman}}, \bibinfo {author} {\bibfnamefont {A.}~\bibnamefont {Mojahed}},\ and\ \bibinfo {author} {\bibfnamefont {M.}~\bibnamefont {Gzal}},\ }\href@noop {} {\bibfield  {journal} {\bibinfo  {journal} {Nonlinear Dynamics}\ }\textbf {\bibinfo {volume} {108}},\ \bibinfo {pages} {711} (\bibinfo {year} {2022})}\BibitemShut {NoStop}%
\bibitem [{\citenamefont {Chen}\ \emph {et~al.}(2017)\citenamefont {Chen}, \citenamefont {Zanette}, \citenamefont {Czaplewski}, \citenamefont {Shaw},\ and\ \citenamefont {L{\'o}pez}}]{chen2017direct}%
  \BibitemOpen
  \bibfield  {author} {\bibinfo {author} {\bibfnamefont {C.}~\bibnamefont {Chen}}, \bibinfo {author} {\bibfnamefont {D.~H.}\ \bibnamefont {Zanette}}, \bibinfo {author} {\bibfnamefont {D.~A.}\ \bibnamefont {Czaplewski}}, \bibinfo {author} {\bibfnamefont {S.}~\bibnamefont {Shaw}},\ and\ \bibinfo {author} {\bibfnamefont {D.}~\bibnamefont {L{\'o}pez}},\ }\href@noop {} {\bibfield  {journal} {\bibinfo  {journal} {Nature Communications}\ }\textbf {\bibinfo {volume} {8}},\ \bibinfo {pages} {1} (\bibinfo {year} {2017})}\BibitemShut {NoStop}%
\bibitem [{\citenamefont {Matheny}\ \emph {et~al.}(2019)\citenamefont {Matheny}, \citenamefont {Emenheiser}, \citenamefont {Fon}, \citenamefont {Chapman}, \citenamefont {Salova}, \citenamefont {Rohden}, \citenamefont {Li}, \citenamefont {Hudoba~de Badyn}, \citenamefont {P{\'o}sfai}, \citenamefont {Duenas-Osorio} \emph {et~al.}}]{matheny2019exotic}%
  \BibitemOpen
  \bibfield  {author} {\bibinfo {author} {\bibfnamefont {M.~H.}\ \bibnamefont {Matheny}}, \bibinfo {author} {\bibfnamefont {J.}~\bibnamefont {Emenheiser}}, \bibinfo {author} {\bibfnamefont {W.}~\bibnamefont {Fon}}, \bibinfo {author} {\bibfnamefont {A.}~\bibnamefont {Chapman}}, \bibinfo {author} {\bibfnamefont {A.}~\bibnamefont {Salova}}, \bibinfo {author} {\bibfnamefont {M.}~\bibnamefont {Rohden}}, \bibinfo {author} {\bibfnamefont {J.}~\bibnamefont {Li}}, \bibinfo {author} {\bibfnamefont {M.}~\bibnamefont {Hudoba~de Badyn}}, \bibinfo {author} {\bibfnamefont {M.}~\bibnamefont {P{\'o}sfai}}, \bibinfo {author} {\bibfnamefont {L.}~\bibnamefont {Duenas-Osorio}}, \emph {et~al.},\ }\href@noop {} {\bibfield  {journal} {\bibinfo  {journal} {Science}\ }\textbf {\bibinfo {volume} {363}},\ \bibinfo {pages} {eaav7932} (\bibinfo {year} {2019})}\BibitemShut {NoStop}%
\bibitem [{\citenamefont {G{\"u}ttinger}\ \emph {et~al.}(2017)\citenamefont {G{\"u}ttinger}, \citenamefont {Noury}, \citenamefont {Weber}, \citenamefont {Eriksson}, \citenamefont {Lagoin}, \citenamefont {Moser}, \citenamefont {Eichler}, \citenamefont {Wallraff}, \citenamefont {Isacsson},\ and\ \citenamefont {Bachtold}}]{guttinger2017energy}%
  \BibitemOpen
  \bibfield  {author} {\bibinfo {author} {\bibfnamefont {J.}~\bibnamefont {G{\"u}ttinger}}, \bibinfo {author} {\bibfnamefont {A.}~\bibnamefont {Noury}}, \bibinfo {author} {\bibfnamefont {P.}~\bibnamefont {Weber}}, \bibinfo {author} {\bibfnamefont {A.~M.}\ \bibnamefont {Eriksson}}, \bibinfo {author} {\bibfnamefont {C.}~\bibnamefont {Lagoin}}, \bibinfo {author} {\bibfnamefont {J.}~\bibnamefont {Moser}}, \bibinfo {author} {\bibfnamefont {C.}~\bibnamefont {Eichler}}, \bibinfo {author} {\bibfnamefont {A.}~\bibnamefont {Wallraff}}, \bibinfo {author} {\bibfnamefont {A.}~\bibnamefont {Isacsson}},\ and\ \bibinfo {author} {\bibfnamefont {A.}~\bibnamefont {Bachtold}},\ }\href@noop {} {\bibfield  {journal} {\bibinfo  {journal} {Nature Nanotechnology}\ }\textbf {\bibinfo {volume} {12}},\ \bibinfo {pages} {631} (\bibinfo {year} {2017})}\BibitemShut {NoStop}%
\bibitem [{\citenamefont {Ke{\c{s}}kekler}\ \emph {et~al.}(2021)\citenamefont {Ke{\c{s}}kekler}, \citenamefont {Shoshani}, \citenamefont {Lee}, \citenamefont {van~der Zant}, \citenamefont {Steeneken},\ and\ \citenamefont {Alijani}}]{Kecskekler2021tuning}%
  \BibitemOpen
  \bibfield  {author} {\bibinfo {author} {\bibfnamefont {A.}~\bibnamefont {Ke{\c{s}}kekler}}, \bibinfo {author} {\bibfnamefont {O.}~\bibnamefont {Shoshani}}, \bibinfo {author} {\bibfnamefont {M.}~\bibnamefont {Lee}}, \bibinfo {author} {\bibfnamefont {H.~S.}\ \bibnamefont {van~der Zant}}, \bibinfo {author} {\bibfnamefont {P.~G.}\ \bibnamefont {Steeneken}},\ and\ \bibinfo {author} {\bibfnamefont {F.}~\bibnamefont {Alijani}},\ }\href@noop {} {\bibfield  {journal} {\bibinfo  {journal} {Nature Communications}\ }\textbf {\bibinfo {volume} {12}},\ \bibinfo {pages} {1} (\bibinfo {year} {2021})}\BibitemShut {NoStop}%
\bibitem [{\citenamefont {Yang}\ \emph {et~al.}(2021)\citenamefont {Yang}, \citenamefont {Hellbach}, \citenamefont {Rochau}, \citenamefont {Belzig}, \citenamefont {Weig}, \citenamefont {Rastelli},\ and\ \citenamefont {Scheer}}]{yang2021persistent}%
  \BibitemOpen
  \bibfield  {author} {\bibinfo {author} {\bibfnamefont {F.}~\bibnamefont {Yang}}, \bibinfo {author} {\bibfnamefont {F.}~\bibnamefont {Hellbach}}, \bibinfo {author} {\bibfnamefont {F.}~\bibnamefont {Rochau}}, \bibinfo {author} {\bibfnamefont {W.}~\bibnamefont {Belzig}}, \bibinfo {author} {\bibfnamefont {E.~M.}\ \bibnamefont {Weig}}, \bibinfo {author} {\bibfnamefont {G.}~\bibnamefont {Rastelli}},\ and\ \bibinfo {author} {\bibfnamefont {E.}~\bibnamefont {Scheer}},\ }\href@noop {} {\bibfield  {journal} {\bibinfo  {journal} {Physical Review Letters}\ }\textbf {\bibinfo {volume} {127}},\ \bibinfo {pages} {014304} (\bibinfo {year} {2021})}\BibitemShut {NoStop}%
\bibitem [{\citenamefont {Houri}\ \emph {et~al.}(2020{\natexlab{a}})\citenamefont {Houri}, \citenamefont {Asano}, \citenamefont {Yamaguchi}, \citenamefont {Yoshimura}, \citenamefont {Koike},\ and\ \citenamefont {Minati}}]{houri2020generic}%
  \BibitemOpen
  \bibfield  {author} {\bibinfo {author} {\bibfnamefont {S.}~\bibnamefont {Houri}}, \bibinfo {author} {\bibfnamefont {M.}~\bibnamefont {Asano}}, \bibinfo {author} {\bibfnamefont {H.}~\bibnamefont {Yamaguchi}}, \bibinfo {author} {\bibfnamefont {N.}~\bibnamefont {Yoshimura}}, \bibinfo {author} {\bibfnamefont {Y.}~\bibnamefont {Koike}},\ and\ \bibinfo {author} {\bibfnamefont {L.}~\bibnamefont {Minati}},\ }\href@noop {} {\bibfield  {journal} {\bibinfo  {journal} {Physical Review Letters}\ }\textbf {\bibinfo {volume} {125}},\ \bibinfo {pages} {174301} (\bibinfo {year} {2020}{\natexlab{a}})}\BibitemShut {NoStop}%
\bibitem [{\citenamefont {Eriksson}\ \emph {et~al.}(2023)\citenamefont {Eriksson}, \citenamefont {Shoshani}, \citenamefont {L{\'o}pez}, \citenamefont {Shaw},\ and\ \citenamefont {Czaplewski}}]{eriksson2023controllable}%
  \BibitemOpen
  \bibfield  {author} {\bibinfo {author} {\bibfnamefont {A.~M.}\ \bibnamefont {Eriksson}}, \bibinfo {author} {\bibfnamefont {O.}~\bibnamefont {Shoshani}}, \bibinfo {author} {\bibfnamefont {D.}~\bibnamefont {L{\'o}pez}}, \bibinfo {author} {\bibfnamefont {S.~W.}\ \bibnamefont {Shaw}},\ and\ \bibinfo {author} {\bibfnamefont {D.~A.}\ \bibnamefont {Czaplewski}},\ }\href@noop {} {\bibfield  {journal} {\bibinfo  {journal} {Nature Communications}\ }\textbf {\bibinfo {volume} {14}},\ \bibinfo {pages} {161} (\bibinfo {year} {2023})}\BibitemShut {NoStop}%
\bibitem [{\citenamefont {Belardinelli}\ \emph {et~al.}(2025)\citenamefont {Belardinelli}, \citenamefont {Yang}, \citenamefont {Bachtold}, \citenamefont {Dykman},\ and\ \citenamefont {Alijani}}]{belardinelli2025hidden}%
  \BibitemOpen
  \bibfield  {author} {\bibinfo {author} {\bibfnamefont {P.}~\bibnamefont {Belardinelli}}, \bibinfo {author} {\bibfnamefont {W.}~\bibnamefont {Yang}}, \bibinfo {author} {\bibfnamefont {A.}~\bibnamefont {Bachtold}}, \bibinfo {author} {\bibfnamefont {M.}~\bibnamefont {Dykman}},\ and\ \bibinfo {author} {\bibfnamefont {F.}~\bibnamefont {Alijani}},\ }\href@noop {} {\bibfield  {journal} {\bibinfo  {journal} {Nano Letters}\ }\textbf {\bibinfo {volume} {25}},\ \bibinfo {pages} {8443} (\bibinfo {year} {2025})}\BibitemShut {NoStop}%
\bibitem [{\citenamefont {Antonio}\ \emph {et~al.}(2012)\citenamefont {Antonio}, \citenamefont {Zanette},\ and\ \citenamefont {L{\'o}pez}}]{antonio2012frequency}%
  \BibitemOpen
  \bibfield  {author} {\bibinfo {author} {\bibfnamefont {D.}~\bibnamefont {Antonio}}, \bibinfo {author} {\bibfnamefont {D.~H.}\ \bibnamefont {Zanette}},\ and\ \bibinfo {author} {\bibfnamefont {D.}~\bibnamefont {L{\'o}pez}},\ }\href@noop {} {\bibfield  {journal} {\bibinfo  {journal} {Nature Communications}\ }\textbf {\bibinfo {volume} {3}},\ \bibinfo {pages} {806} (\bibinfo {year} {2012})}\BibitemShut {NoStop}%
\bibitem [{\citenamefont {Shoshani}\ \emph {et~al.}(2024)\citenamefont {Shoshani}, \citenamefont {Strachan}, \citenamefont {Czaplewski}, \citenamefont {Lopez},\ and\ \citenamefont {Shaw}}]{shoshani2024extraordinary}%
  \BibitemOpen
  \bibfield  {author} {\bibinfo {author} {\bibfnamefont {O.}~\bibnamefont {Shoshani}}, \bibinfo {author} {\bibfnamefont {S.}~\bibnamefont {Strachan}}, \bibinfo {author} {\bibfnamefont {D.}~\bibnamefont {Czaplewski}}, \bibinfo {author} {\bibfnamefont {D.}~\bibnamefont {Lopez}},\ and\ \bibinfo {author} {\bibfnamefont {S.~W.}\ \bibnamefont {Shaw}},\ }\href@noop {} {\bibfield  {journal} {\bibinfo  {journal} {Physical Review Applied}\ }\textbf {\bibinfo {volume} {22}},\ \bibinfo {pages} {054055} (\bibinfo {year} {2024})}\BibitemShut {NoStop}%
\bibitem [{\citenamefont {Chen}\ and\ \citenamefont {Fan}(2023)}]{chen2023internal}%
  \BibitemOpen
  \bibfield  {author} {\bibinfo {author} {\bibfnamefont {L.-Q.}\ \bibnamefont {Chen}}\ and\ \bibinfo {author} {\bibfnamefont {Y.}~\bibnamefont {Fan}},\ }\href@noop {} {\bibfield  {journal} {\bibinfo  {journal} {Nonlinear Dynamics}\ }\textbf {\bibinfo {volume} {111}},\ \bibinfo {pages} {11703} (\bibinfo {year} {2023})}\BibitemShut {NoStop}%
\bibitem [{\citenamefont {Asadi}\ \emph {et~al.}(2018)\citenamefont {Asadi}, \citenamefont {Yu},\ and\ \citenamefont {Cho}}]{asadi2018nonlinear}%
  \BibitemOpen
  \bibfield  {author} {\bibinfo {author} {\bibfnamefont {K.}~\bibnamefont {Asadi}}, \bibinfo {author} {\bibfnamefont {J.}~\bibnamefont {Yu}},\ and\ \bibinfo {author} {\bibfnamefont {H.}~\bibnamefont {Cho}},\ }\href@noop {} {\bibfield  {journal} {\bibinfo  {journal} {Philosophical Transactions of the Royal Society A: Mathematical, Physical and Engineering Sciences}\ }\textbf {\bibinfo {volume} {376}},\ \bibinfo {pages} {20170141} (\bibinfo {year} {2018})}\BibitemShut {NoStop}%
\bibitem [{\citenamefont {Ke{\c{s}}kekler}\ \emph {et~al.}(2022)\citenamefont {Ke{\c{s}}kekler}, \citenamefont {Arjmandi-Tash}, \citenamefont {Steeneken},\ and\ \citenamefont {Alijani}}]{keskekler2022symmetry}%
  \BibitemOpen
  \bibfield  {author} {\bibinfo {author} {\bibfnamefont {A.}~\bibnamefont {Ke{\c{s}}kekler}}, \bibinfo {author} {\bibfnamefont {H.}~\bibnamefont {Arjmandi-Tash}}, \bibinfo {author} {\bibfnamefont {P.~G.}\ \bibnamefont {Steeneken}},\ and\ \bibinfo {author} {\bibfnamefont {F.}~\bibnamefont {Alijani}},\ }\href@noop {} {\bibfield  {journal} {\bibinfo  {journal} {Nano letters}\ }\textbf {\bibinfo {volume} {22}},\ \bibinfo {pages} {6048} (\bibinfo {year} {2022})}\BibitemShut {NoStop}%
\bibitem [{\citenamefont {Sun}\ \emph {et~al.}(2023)\citenamefont {Sun}, \citenamefont {Yu}, \citenamefont {Zhang}, \citenamefont {Chen}, \citenamefont {Zhou}, \citenamefont {Zhao}, \citenamefont {Gerrard}, \citenamefont {Kwon}, \citenamefont {Vukasin}, \citenamefont {Xiao} \emph {et~al.}}]{sun2023generation}%
  \BibitemOpen
  \bibfield  {author} {\bibinfo {author} {\bibfnamefont {J.}~\bibnamefont {Sun}}, \bibinfo {author} {\bibfnamefont {S.}~\bibnamefont {Yu}}, \bibinfo {author} {\bibfnamefont {H.}~\bibnamefont {Zhang}}, \bibinfo {author} {\bibfnamefont {D.}~\bibnamefont {Chen}}, \bibinfo {author} {\bibfnamefont {X.}~\bibnamefont {Zhou}}, \bibinfo {author} {\bibfnamefont {C.}~\bibnamefont {Zhao}}, \bibinfo {author} {\bibfnamefont {D.~D.}\ \bibnamefont {Gerrard}}, \bibinfo {author} {\bibfnamefont {R.}~\bibnamefont {Kwon}}, \bibinfo {author} {\bibfnamefont {G.}~\bibnamefont {Vukasin}}, \bibinfo {author} {\bibfnamefont {D.}~\bibnamefont {Xiao}}, \emph {et~al.},\ }\href@noop {} {\bibfield  {journal} {\bibinfo  {journal} {Physical Review Applied}\ }\textbf {\bibinfo {volume} {19}},\ \bibinfo {pages} {014031} (\bibinfo {year} {2023})}\BibitemShut {NoStop}%
\bibitem [{\citenamefont {Fu}\ \emph {et~al.}(2025)\citenamefont {Fu}, \citenamefont {Ameye}, \citenamefont {Yang}, \citenamefont {Ko{\v{s}}ata}, \citenamefont {del Pino}, \citenamefont {Zilberberg},\ and\ \citenamefont {Scheer}}]{fu2025sideband}%
  \BibitemOpen
  \bibfield  {author} {\bibinfo {author} {\bibfnamefont {M.}~\bibnamefont {Fu}}, \bibinfo {author} {\bibfnamefont {O.}~\bibnamefont {Ameye}}, \bibinfo {author} {\bibfnamefont {F.}~\bibnamefont {Yang}}, \bibinfo {author} {\bibfnamefont {J.}~\bibnamefont {Ko{\v{s}}ata}}, \bibinfo {author} {\bibfnamefont {J.}~\bibnamefont {del Pino}}, \bibinfo {author} {\bibfnamefont {O.}~\bibnamefont {Zilberberg}},\ and\ \bibinfo {author} {\bibfnamefont {E.}~\bibnamefont {Scheer}},\ }\href@noop {} {\bibfield  {journal} {\bibinfo  {journal} {Physical Review Research}\ }\textbf {\bibinfo {volume} {7}},\ \bibinfo {pages} {033127} (\bibinfo {year} {2025})}\BibitemShut {NoStop}%
\bibitem [{\citenamefont {Wu}\ \emph {et~al.}(2025)\citenamefont {Wu}, \citenamefont {Song}, \citenamefont {Zang}, \citenamefont {Mao}, \citenamefont {Zhang},\ and\ \citenamefont {Shao}}]{wu2025self}%
  \BibitemOpen
  \bibfield  {author} {\bibinfo {author} {\bibfnamefont {J.}~\bibnamefont {Wu}}, \bibinfo {author} {\bibfnamefont {P.}~\bibnamefont {Song}}, \bibinfo {author} {\bibfnamefont {S.}~\bibnamefont {Zang}}, \bibinfo {author} {\bibfnamefont {Z.}~\bibnamefont {Mao}}, \bibinfo {author} {\bibfnamefont {W.}~\bibnamefont {Zhang}},\ and\ \bibinfo {author} {\bibfnamefont {L.}~\bibnamefont {Shao}},\ }\href@noop {} {\bibfield  {journal} {\bibinfo  {journal} {Physical Review Letters}\ }\textbf {\bibinfo {volume} {134}},\ \bibinfo {pages} {107201} (\bibinfo {year} {2025})}\BibitemShut {NoStop}%
\bibitem [{\citenamefont {Yang}\ \emph {et~al.}(2015)\citenamefont {Yang}, \citenamefont {Ng}, \citenamefont {Polunin}, \citenamefont {Chen}, \citenamefont {Strachan}, \citenamefont {Hong}, \citenamefont {Ahn}, \citenamefont {Shoshani}, \citenamefont {Shaw}, \citenamefont {Dykman},\ and\ \citenamefont {Kenny}}]{yang2015experimental}%
  \BibitemOpen
  \bibfield  {author} {\bibinfo {author} {\bibfnamefont {Y.}~\bibnamefont {Yang}}, \bibinfo {author} {\bibfnamefont {E.}~\bibnamefont {Ng}}, \bibinfo {author} {\bibfnamefont {P.}~\bibnamefont {Polunin}}, \bibinfo {author} {\bibfnamefont {Y.}~\bibnamefont {Chen}}, \bibinfo {author} {\bibfnamefont {S.}~\bibnamefont {Strachan}}, \bibinfo {author} {\bibfnamefont {V.}~\bibnamefont {Hong}}, \bibinfo {author} {\bibfnamefont {C.~H.}\ \bibnamefont {Ahn}}, \bibinfo {author} {\bibfnamefont {O.}~\bibnamefont {Shoshani}}, \bibinfo {author} {\bibfnamefont {S.}~\bibnamefont {Shaw}}, \bibinfo {author} {\bibfnamefont {M.}~\bibnamefont {Dykman}},\ and\ \bibinfo {author} {\bibfnamefont {T.}~\bibnamefont {Kenny}},\ }in\ \href@noop {} {\emph {\bibinfo {booktitle} {2015 28th IEEE International Conference on Micro Electro Mechanical Systems (MEMS)}}}\ (\bibinfo {organization} {IEEE},\ \bibinfo {year} {2015})\ pp.\ \bibinfo {pages} {1008--1011}\BibitemShut {NoStop}%
\bibitem [{\citenamefont {Qiao}\ \emph {et~al.}(2023)\citenamefont {Qiao}, \citenamefont {Shi}, \citenamefont {Xu}, \citenamefont {Wei}, \citenamefont {Elhady}, \citenamefont {Abdel-Rahman}, \citenamefont {Huan},\ and\ \citenamefont {Zhang}}]{qiao2023frequency}%
  \BibitemOpen
  \bibfield  {author} {\bibinfo {author} {\bibfnamefont {Y.}~\bibnamefont {Qiao}}, \bibinfo {author} {\bibfnamefont {Z.}~\bibnamefont {Shi}}, \bibinfo {author} {\bibfnamefont {Y.}~\bibnamefont {Xu}}, \bibinfo {author} {\bibfnamefont {X.}~\bibnamefont {Wei}}, \bibinfo {author} {\bibfnamefont {A.}~\bibnamefont {Elhady}}, \bibinfo {author} {\bibfnamefont {E.}~\bibnamefont {Abdel-Rahman}}, \bibinfo {author} {\bibfnamefont {R.}~\bibnamefont {Huan}},\ and\ \bibinfo {author} {\bibfnamefont {W.}~\bibnamefont {Zhang}},\ }\href@noop {} {\bibfield  {journal} {\bibinfo  {journal} {Microsystems \& Nanoengineering}\ }\textbf {\bibinfo {volume} {9}},\ \bibinfo {pages} {58} (\bibinfo {year} {2023})}\BibitemShut {NoStop}%
\bibitem [{\citenamefont {Fan}\ \emph {et~al.}(2024)\citenamefont {Fan}, \citenamefont {Zhang}, \citenamefont {Niu},\ and\ \citenamefont {Chen}}]{fan2024internal}%
  \BibitemOpen
  \bibfield  {author} {\bibinfo {author} {\bibfnamefont {Y.}~\bibnamefont {Fan}}, \bibinfo {author} {\bibfnamefont {Y.}~\bibnamefont {Zhang}}, \bibinfo {author} {\bibfnamefont {M.-Q.}\ \bibnamefont {Niu}},\ and\ \bibinfo {author} {\bibfnamefont {L.-Q.}\ \bibnamefont {Chen}},\ }\href@noop {} {\bibfield  {journal} {\bibinfo  {journal} {Mechanical Systems and Signal Processing}\ }\textbf {\bibinfo {volume} {211}},\ \bibinfo {pages} {111176} (\bibinfo {year} {2024})}\BibitemShut {NoStop}%
\bibitem [{\citenamefont {Monteil}\ \emph {et~al.}(2014)\citenamefont {Monteil}, \citenamefont {Touz{\'e}}, \citenamefont {Thomas},\ and\ \citenamefont {Benacchio}}]{monteil2014nonlinear}%
  \BibitemOpen
  \bibfield  {author} {\bibinfo {author} {\bibfnamefont {M.}~\bibnamefont {Monteil}}, \bibinfo {author} {\bibfnamefont {C.}~\bibnamefont {Touz{\'e}}}, \bibinfo {author} {\bibfnamefont {O.}~\bibnamefont {Thomas}},\ and\ \bibinfo {author} {\bibfnamefont {S.}~\bibnamefont {Benacchio}},\ }\href@noop {} {\bibfield  {journal} {\bibinfo  {journal} {Nonlinear Dynamics}\ }\textbf {\bibinfo {volume} {75}},\ \bibinfo {pages} {175} (\bibinfo {year} {2014})}\BibitemShut {NoStop}%
\bibitem [{\citenamefont {Li}\ \emph {et~al.}(2023)\citenamefont {Li}, \citenamefont {Xu}, \citenamefont {Norte}, \citenamefont {Arag{\'o}n}, \citenamefont {Van~Keulen}, \citenamefont {Alijani},\ and\ \citenamefont {Steeneken}}]{li2023tuning}%
  \BibitemOpen
  \bibfield  {author} {\bibinfo {author} {\bibfnamefont {Z.}~\bibnamefont {Li}}, \bibinfo {author} {\bibfnamefont {M.}~\bibnamefont {Xu}}, \bibinfo {author} {\bibfnamefont {R.~A.}\ \bibnamefont {Norte}}, \bibinfo {author} {\bibfnamefont {A.~M.}\ \bibnamefont {Arag{\'o}n}}, \bibinfo {author} {\bibfnamefont {F.}~\bibnamefont {Van~Keulen}}, \bibinfo {author} {\bibfnamefont {F.}~\bibnamefont {Alijani}},\ and\ \bibinfo {author} {\bibfnamefont {P.~G.}\ \bibnamefont {Steeneken}},\ }\href@noop {} {\bibfield  {journal} {\bibinfo  {journal} {Applied Physics Letters}\ }\textbf {\bibinfo {volume} {122}},\ \bibinfo {pages} {013501} (\bibinfo {year} {2023})}\BibitemShut {NoStop}%
\bibitem [{\citenamefont {Shin}\ \emph {et~al.}(2022)\citenamefont {Shin}, \citenamefont {Cupertino}, \citenamefont {de~Jong}, \citenamefont {Steeneken}, \citenamefont {Bessa},\ and\ \citenamefont {Norte}}]{Spiderweb}%
  \BibitemOpen
  \bibfield  {author} {\bibinfo {author} {\bibfnamefont {D.}~\bibnamefont {Shin}}, \bibinfo {author} {\bibfnamefont {A.}~\bibnamefont {Cupertino}}, \bibinfo {author} {\bibfnamefont {M.~H.}\ \bibnamefont {de~Jong}}, \bibinfo {author} {\bibfnamefont {P.~G.}\ \bibnamefont {Steeneken}}, \bibinfo {author} {\bibfnamefont {M.~A.}\ \bibnamefont {Bessa}},\ and\ \bibinfo {author} {\bibfnamefont {R.~A.}\ \bibnamefont {Norte}},\ }\href@noop {} {\bibfield  {journal} {\bibinfo  {journal} {Advanced Materials}\ ,\ \bibinfo {pages} {2106248}} (\bibinfo {year} {2022})}\BibitemShut {NoStop}%
\bibitem [{\citenamefont {Fedorov}\ \emph {et~al.}(2020)\citenamefont {Fedorov}, \citenamefont {Beccari}, \citenamefont {Engelsen},\ and\ \citenamefont {Kippenberg}}]{fedorov2020fractal}%
  \BibitemOpen
  \bibfield  {author} {\bibinfo {author} {\bibfnamefont {S.~A.}\ \bibnamefont {Fedorov}}, \bibinfo {author} {\bibfnamefont {A.}~\bibnamefont {Beccari}}, \bibinfo {author} {\bibfnamefont {N.~J.}\ \bibnamefont {Engelsen}},\ and\ \bibinfo {author} {\bibfnamefont {T.~J.}\ \bibnamefont {Kippenberg}},\ }\href@noop {} {\bibfield  {journal} {\bibinfo  {journal} {Physical Review Letters}\ }\textbf {\bibinfo {volume} {124}},\ \bibinfo {pages} {025502} (\bibinfo {year} {2020})}\BibitemShut {NoStop}%
\bibitem [{\citenamefont {Li}\ \emph {et~al.}(2024)\citenamefont {Li}, \citenamefont {Xu}, \citenamefont {Norte}, \citenamefont {Arag{\'o}n}, \citenamefont {Steeneken},\ and\ \citenamefont {Alijani}}]{li2024strain}%
  \BibitemOpen
  \bibfield  {author} {\bibinfo {author} {\bibfnamefont {Z.}~\bibnamefont {Li}}, \bibinfo {author} {\bibfnamefont {M.}~\bibnamefont {Xu}}, \bibinfo {author} {\bibfnamefont {R.~A.}\ \bibnamefont {Norte}}, \bibinfo {author} {\bibfnamefont {A.~M.}\ \bibnamefont {Arag{\'o}n}}, \bibinfo {author} {\bibfnamefont {P.~G.}\ \bibnamefont {Steeneken}},\ and\ \bibinfo {author} {\bibfnamefont {F.}~\bibnamefont {Alijani}},\ }\href@noop {} {\bibfield  {journal} {\bibinfo  {journal} {Communications Physics}\ }\textbf {\bibinfo {volume} {7}},\ \bibinfo {pages} {53} (\bibinfo {year} {2024})}\BibitemShut {NoStop}%
\bibitem [{sup()}]{supplemental}%
  \BibitemOpen
  \href@noop {} {\bibinfo {title} {See {S}upplemental {M}aterial at \url{https://journals.aps.org/supplemental/10.1103/73wb-22pz} for the details of the fabrication, measurement, modeling of mode coupling, and numerical simulation results, which includes {R}efs.~\cite{xu2024high,villanueva2014evidence,hauer2013general,li2024strain,miller2018effective,nayfeh2008nonlinear,li2023tuning,yang2015experimental}}}\BibitemShut {NoStop}%
\bibitem [{\citenamefont {Xu}\ \emph {et~al.}(2024)\citenamefont {Xu}, \citenamefont {Shin}, \citenamefont {Sberna}, \citenamefont {van~der Kolk}, \citenamefont {Cupertino}, \citenamefont {Bessa},\ and\ \citenamefont {Norte}}]{xu2024high}%
  \BibitemOpen
  \bibfield  {author} {\bibinfo {author} {\bibfnamefont {M.}~\bibnamefont {Xu}}, \bibinfo {author} {\bibfnamefont {D.}~\bibnamefont {Shin}}, \bibinfo {author} {\bibfnamefont {P.~M.}\ \bibnamefont {Sberna}}, \bibinfo {author} {\bibfnamefont {R.}~\bibnamefont {van~der Kolk}}, \bibinfo {author} {\bibfnamefont {A.}~\bibnamefont {Cupertino}}, \bibinfo {author} {\bibfnamefont {M.~A.}\ \bibnamefont {Bessa}},\ and\ \bibinfo {author} {\bibfnamefont {R.~A.}\ \bibnamefont {Norte}},\ }\href@noop {} {\bibfield  {journal} {\bibinfo  {journal} {Advanced Materials}\ }\textbf {\bibinfo {volume} {36}},\ \bibinfo {pages} {2306513} (\bibinfo {year} {2024})}\BibitemShut {NoStop}%
\bibitem [{\citenamefont {Villanueva}\ and\ \citenamefont {Schmid}(2014)}]{villanueva2014evidence}%
  \BibitemOpen
  \bibfield  {author} {\bibinfo {author} {\bibfnamefont {L.~G.}\ \bibnamefont {Villanueva}}\ and\ \bibinfo {author} {\bibfnamefont {S.}~\bibnamefont {Schmid}},\ }\href@noop {} {\bibfield  {journal} {\bibinfo  {journal} {Physical Review Letters}\ }\textbf {\bibinfo {volume} {113}},\ \bibinfo {pages} {227201} (\bibinfo {year} {2014})}\BibitemShut {NoStop}%
\bibitem [{\citenamefont {Hauer}\ \emph {et~al.}(2013)\citenamefont {Hauer}, \citenamefont {Doolin}, \citenamefont {Beach},\ and\ \citenamefont {Davis}}]{hauer2013general}%
  \BibitemOpen
  \bibfield  {author} {\bibinfo {author} {\bibfnamefont {B.}~\bibnamefont {Hauer}}, \bibinfo {author} {\bibfnamefont {C.}~\bibnamefont {Doolin}}, \bibinfo {author} {\bibfnamefont {K.}~\bibnamefont {Beach}},\ and\ \bibinfo {author} {\bibfnamefont {J.}~\bibnamefont {Davis}},\ }\href@noop {} {\bibfield  {journal} {\bibinfo  {journal} {Annals of Physics}\ }\textbf {\bibinfo {volume} {339}},\ \bibinfo {pages} {181} (\bibinfo {year} {2013})}\BibitemShut {NoStop}%
\bibitem [{\citenamefont {Miller}\ \emph {et~al.}(2018)\citenamefont {Miller}, \citenamefont {Ansari}, \citenamefont {Heinz}, \citenamefont {Chen}, \citenamefont {Flader}, \citenamefont {Shin}, \citenamefont {Villanueva},\ and\ \citenamefont {Kenny}}]{miller2018effective}%
  \BibitemOpen
  \bibfield  {author} {\bibinfo {author} {\bibfnamefont {J.~M.~L.}\ \bibnamefont {Miller}}, \bibinfo {author} {\bibfnamefont {A.}~\bibnamefont {Ansari}}, \bibinfo {author} {\bibfnamefont {D.~B.}\ \bibnamefont {Heinz}}, \bibinfo {author} {\bibfnamefont {Y.}~\bibnamefont {Chen}}, \bibinfo {author} {\bibfnamefont {I.~B.}\ \bibnamefont {Flader}}, \bibinfo {author} {\bibfnamefont {D.~D.}\ \bibnamefont {Shin}}, \bibinfo {author} {\bibfnamefont {L.~G.}\ \bibnamefont {Villanueva}},\ and\ \bibinfo {author} {\bibfnamefont {T.~W.}\ \bibnamefont {Kenny}},\ }\href@noop {} {\bibfield  {journal} {\bibinfo  {journal} {Applied Physics Reviews}\ }\textbf {\bibinfo {volume} {5}} (\bibinfo {year} {2018})}\BibitemShut {NoStop}%
\bibitem [{\citenamefont {Nayfeh}\ and\ \citenamefont {Mook}(2008)}]{nayfeh2008nonlinear}%
  \BibitemOpen
  \bibfield  {author} {\bibinfo {author} {\bibfnamefont {A.~H.}\ \bibnamefont {Nayfeh}}\ and\ \bibinfo {author} {\bibfnamefont {D.~T.}\ \bibnamefont {Mook}},\ }\href@noop {} {\emph {\bibinfo {title} {Nonlinear oscillations}}}\ (\bibinfo  {publisher} {John Wiley \& Sons},\ \bibinfo {year} {2008})\BibitemShut {NoStop}%
\bibitem [{\citenamefont {Li}\ \emph {et~al.}(2025)\citenamefont {Li}, \citenamefont {Alijani}, \citenamefont {Sarafraz}, \citenamefont {Xu}, \citenamefont {Norte}, \citenamefont {Arag{\'o}n},\ and\ \citenamefont {Steeneken}}]{li2025finite}%
  \BibitemOpen
  \bibfield  {author} {\bibinfo {author} {\bibfnamefont {Z.}~\bibnamefont {Li}}, \bibinfo {author} {\bibfnamefont {F.}~\bibnamefont {Alijani}}, \bibinfo {author} {\bibfnamefont {A.}~\bibnamefont {Sarafraz}}, \bibinfo {author} {\bibfnamefont {M.}~\bibnamefont {Xu}}, \bibinfo {author} {\bibfnamefont {R.~A.}\ \bibnamefont {Norte}}, \bibinfo {author} {\bibfnamefont {A.~M.}\ \bibnamefont {Arag{\'o}n}},\ and\ \bibinfo {author} {\bibfnamefont {P.~G.}\ \bibnamefont {Steeneken}},\ }\href@noop {} {\bibfield  {journal} {\bibinfo  {journal} {Microsystems \& Nanoengineering}\ }\textbf {\bibinfo {volume} {11}},\ \bibinfo {pages} {16} (\bibinfo {year} {2025})}\BibitemShut {NoStop}%
\bibitem [{\citenamefont {Schmid}\ \emph {et~al.}(2016)\citenamefont {Schmid}, \citenamefont {Villanueva},\ and\ \citenamefont {Roukes}}]{schmid2016fundamentals}%
  \BibitemOpen
  \bibfield  {author} {\bibinfo {author} {\bibfnamefont {S.}~\bibnamefont {Schmid}}, \bibinfo {author} {\bibfnamefont {L.~G.}\ \bibnamefont {Villanueva}},\ and\ \bibinfo {author} {\bibfnamefont {M.~L.}\ \bibnamefont {Roukes}},\ }\href@noop {} {\emph {\bibinfo {title} {Fundamentals of nanomechanical resonators}}},\ Vol.~\bibinfo {volume} {49}\ (\bibinfo  {publisher} {Springer},\ \bibinfo {year} {2016})\BibitemShut {NoStop}%
\bibitem [{\citenamefont {Houri}\ \emph {et~al.}(2020{\natexlab{b}})\citenamefont {Houri}, \citenamefont {Hatanaka}, \citenamefont {Asano},\ and\ \citenamefont {Yamaguchi}}]{houri2020demonstration}%
  \BibitemOpen
  \bibfield  {author} {\bibinfo {author} {\bibfnamefont {S.}~\bibnamefont {Houri}}, \bibinfo {author} {\bibfnamefont {D.}~\bibnamefont {Hatanaka}}, \bibinfo {author} {\bibfnamefont {M.}~\bibnamefont {Asano}},\ and\ \bibinfo {author} {\bibfnamefont {H.}~\bibnamefont {Yamaguchi}},\ }\href@noop {} {\bibfield  {journal} {\bibinfo  {journal} {Physical Review Applied}\ }\textbf {\bibinfo {volume} {13}},\ \bibinfo {pages} {014049} (\bibinfo {year} {2020}{\natexlab{b}})}\BibitemShut {NoStop}%
\bibitem [{\citenamefont {Muravyov}\ and\ \citenamefont {Rizzi}(2003)}]{muravyov2003determination}%
  \BibitemOpen
  \bibfield  {author} {\bibinfo {author} {\bibfnamefont {A.~A.}\ \bibnamefont {Muravyov}}\ and\ \bibinfo {author} {\bibfnamefont {S.~A.}\ \bibnamefont {Rizzi}},\ }\href@noop {} {\bibfield  {journal} {\bibinfo  {journal} {Computers \& Structures}\ }\textbf {\bibinfo {volume} {81}},\ \bibinfo {pages} {1513} (\bibinfo {year} {2003})}\BibitemShut {NoStop}%
\bibitem [{\citenamefont {Ke{\c{s}}kekler}\ \emph {et~al.}(2023)\citenamefont {Ke{\c{s}}kekler}, \citenamefont {Bos}, \citenamefont {Arag{\'o}n}, \citenamefont {Steeneken},\ and\ \citenamefont {Alijani}}]{kecskekler2023multimode}%
  \BibitemOpen
  \bibfield  {author} {\bibinfo {author} {\bibfnamefont {A.}~\bibnamefont {Ke{\c{s}}kekler}}, \bibinfo {author} {\bibfnamefont {V.}~\bibnamefont {Bos}}, \bibinfo {author} {\bibfnamefont {A.~M.}\ \bibnamefont {Arag{\'o}n}}, \bibinfo {author} {\bibfnamefont {P.~G.}\ \bibnamefont {Steeneken}},\ and\ \bibinfo {author} {\bibfnamefont {F.}~\bibnamefont {Alijani}},\ }\href@noop {} {\bibfield  {journal} {\bibinfo  {journal} {Physical Review Applied}\ }\textbf {\bibinfo {volume} {20}},\ \bibinfo {pages} {064020} (\bibinfo {year} {2023})}\BibitemShut {NoStop}%
\bibitem [{\citenamefont {Dhooge}\ \emph {et~al.}(2008)\citenamefont {Dhooge}, \citenamefont {Govaerts}, \citenamefont {Kuznetsov}, \citenamefont {Meijer},\ and\ \citenamefont {Sautois}}]{Matcont}%
  \BibitemOpen
  \bibfield  {author} {\bibinfo {author} {\bibfnamefont {A.}~\bibnamefont {Dhooge}}, \bibinfo {author} {\bibfnamefont {W.}~\bibnamefont {Govaerts}}, \bibinfo {author} {\bibfnamefont {Y.~A.}\ \bibnamefont {Kuznetsov}}, \bibinfo {author} {\bibfnamefont {H.~G.~E.}\ \bibnamefont {Meijer}},\ and\ \bibinfo {author} {\bibfnamefont {B.}~\bibnamefont {Sautois}},\ }\href@noop {} {\bibfield  {journal} {\bibinfo  {journal} {Mathematical and Computer Modelling of Dynamical Systems}\ }\textbf {\bibinfo {volume} {14}},\ \bibinfo {pages} {147} (\bibinfo {year} {2008})}\BibitemShut {NoStop}%
\bibitem [{\citenamefont {Hellbach}\ \emph {et~al.}(2024)\citenamefont {Hellbach}, \citenamefont {De~Bernardis}, \citenamefont {Saur}, \citenamefont {Carusotto}, \citenamefont {Belzig},\ and\ \citenamefont {Rastelli}}]{hellbach2024nonlinearity}%
  \BibitemOpen
  \bibfield  {author} {\bibinfo {author} {\bibfnamefont {F.}~\bibnamefont {Hellbach}}, \bibinfo {author} {\bibfnamefont {D.}~\bibnamefont {De~Bernardis}}, \bibinfo {author} {\bibfnamefont {M.}~\bibnamefont {Saur}}, \bibinfo {author} {\bibfnamefont {I.}~\bibnamefont {Carusotto}}, \bibinfo {author} {\bibfnamefont {W.}~\bibnamefont {Belzig}},\ and\ \bibinfo {author} {\bibfnamefont {G.}~\bibnamefont {Rastelli}},\ }\href@noop {} {\bibfield  {journal} {\bibinfo  {journal} {New Journal of Physics}\ }\textbf {\bibinfo {volume} {26}},\ \bibinfo {pages} {103020} (\bibinfo {year} {2024})}\BibitemShut {NoStop}%
\bibitem [{\citenamefont {Defoort}\ \emph {et~al.}(2022)\citenamefont {Defoort}, \citenamefont {Hentz}, \citenamefont {Shaw},\ and\ \citenamefont {Shoshani}}]{defoort2022amplitude}%
  \BibitemOpen
  \bibfield  {author} {\bibinfo {author} {\bibfnamefont {M.}~\bibnamefont {Defoort}}, \bibinfo {author} {\bibfnamefont {S.}~\bibnamefont {Hentz}}, \bibinfo {author} {\bibfnamefont {S.~W.}\ \bibnamefont {Shaw}},\ and\ \bibinfo {author} {\bibfnamefont {O.}~\bibnamefont {Shoshani}},\ }\href@noop {} {\bibfield  {journal} {\bibinfo  {journal} {Communications Physics}\ }\textbf {\bibinfo {volume} {5}},\ \bibinfo {pages} {93} (\bibinfo {year} {2022})}\BibitemShut {NoStop}%
\bibitem [{\citenamefont {Villanueva}\ \emph {et~al.}(2011)\citenamefont {Villanueva}, \citenamefont {Karabalin}, \citenamefont {Matheny}, \citenamefont {Kenig}, \citenamefont {Cross},\ and\ \citenamefont {Roukes}}]{villanueva2011nanoscale}%
  \BibitemOpen
  \bibfield  {author} {\bibinfo {author} {\bibfnamefont {L.~G.}\ \bibnamefont {Villanueva}}, \bibinfo {author} {\bibfnamefont {R.~B.}\ \bibnamefont {Karabalin}}, \bibinfo {author} {\bibfnamefont {M.~H.}\ \bibnamefont {Matheny}}, \bibinfo {author} {\bibfnamefont {E.}~\bibnamefont {Kenig}}, \bibinfo {author} {\bibfnamefont {M.~C.}\ \bibnamefont {Cross}},\ and\ \bibinfo {author} {\bibfnamefont {M.~L.}\ \bibnamefont {Roukes}},\ }\href@noop {} {\bibfield  {journal} {\bibinfo  {journal} {Nano Letters}\ }\textbf {\bibinfo {volume} {11}},\ \bibinfo {pages} {5054} (\bibinfo {year} {2011})}\BibitemShut {NoStop}%
\bibitem [{\citenamefont {Jin}\ \emph {et~al.}(2025)\citenamefont {Jin}, \citenamefont {Baker}, \citenamefont {Romero}, \citenamefont {Arora}, \citenamefont {Mauranyapin}, \citenamefont {Hirsch}, \citenamefont {Harris},\ and\ \citenamefont {Bowen}}]{jin2025nanomechanical}%
  \BibitemOpen
  \bibfield  {author} {\bibinfo {author} {\bibfnamefont {X.}~\bibnamefont {Jin}}, \bibinfo {author} {\bibfnamefont {C.~G.}\ \bibnamefont {Baker}}, \bibinfo {author} {\bibfnamefont {E.}~\bibnamefont {Romero}}, \bibinfo {author} {\bibfnamefont {N.}~\bibnamefont {Arora}}, \bibinfo {author} {\bibfnamefont {N.~P.}\ \bibnamefont {Mauranyapin}}, \bibinfo {author} {\bibfnamefont {T.~M.}\ \bibnamefont {Hirsch}}, \bibinfo {author} {\bibfnamefont {G.~I.}\ \bibnamefont {Harris}},\ and\ \bibinfo {author} {\bibfnamefont {W.~P.}\ \bibnamefont {Bowen}},\ }\href@noop {} {\bibfield  {journal} {\bibinfo  {journal} {arXiv preprint arXiv:2509.11560}\ } (\bibinfo {year} {2025})}\BibitemShut {NoStop}%
\bibitem [{\citenamefont {Mahboob}\ and\ \citenamefont {Yamaguchi}(2008)}]{mahboob2008bit}%
  \BibitemOpen
  \bibfield  {author} {\bibinfo {author} {\bibfnamefont {I.}~\bibnamefont {Mahboob}}\ and\ \bibinfo {author} {\bibfnamefont {H.}~\bibnamefont {Yamaguchi}},\ }\href@noop {} {\bibfield  {journal} {\bibinfo  {journal} {Nature Nanotechnology}\ }\textbf {\bibinfo {volume} {3}},\ \bibinfo {pages} {275} (\bibinfo {year} {2008})}\BibitemShut {NoStop}%
\bibitem [{\citenamefont {Bagheri}\ \emph {et~al.}(2011)\citenamefont {Bagheri}, \citenamefont {Poot}, \citenamefont {Li}, \citenamefont {Pernice},\ and\ \citenamefont {Tang}}]{bagheri2011dynamic}%
  \BibitemOpen
  \bibfield  {author} {\bibinfo {author} {\bibfnamefont {M.}~\bibnamefont {Bagheri}}, \bibinfo {author} {\bibfnamefont {M.}~\bibnamefont {Poot}}, \bibinfo {author} {\bibfnamefont {M.}~\bibnamefont {Li}}, \bibinfo {author} {\bibfnamefont {W.~P.}\ \bibnamefont {Pernice}},\ and\ \bibinfo {author} {\bibfnamefont {H.~X.}\ \bibnamefont {Tang}},\ }\href@noop {} {\bibfield  {journal} {\bibinfo  {journal} {Nature Nanotechnology}\ }\textbf {\bibinfo {volume} {6}},\ \bibinfo {pages} {726} (\bibinfo {year} {2011})}\BibitemShut {NoStop}%
\bibitem [{\citenamefont {Yao}\ and\ \citenamefont {Hikihara}(2014)}]{yao2014logic}%
  \BibitemOpen
  \bibfield  {author} {\bibinfo {author} {\bibfnamefont {A.}~\bibnamefont {Yao}}\ and\ \bibinfo {author} {\bibfnamefont {T.}~\bibnamefont {Hikihara}},\ }\href@noop {} {\bibfield  {journal} {\bibinfo  {journal} {Applied Physics Letters}\ }\textbf {\bibinfo {volume} {105}} (\bibinfo {year} {2014})}\BibitemShut {NoStop}%
\end{thebibliography}%

\clearpage



\section*{Supplementary material}

# Supplementary Information: Cascade of Modal Interactions in Nanomechanical Resonators with Soft Clamping

Zichao Li[*,1], Minxing Xu[1,2], Richard A. Norte[1,2], Alejandro M. Aragón[1], Peter G. Steeneken[1,2], and Farbod Alijani[*,1]

[1]Department of Precision and Microsystems Engineering, Delft University of Technology, Mekelweg 2, 2628 CD Delft, The Netherlands

[2]Kavli Institute of Nanoscience, Delft University of Technology, Lorentzweg 1, 2628 CJ Delft, The Netherlands

June 25, 2025

## S1. Fabrication process and material properties

The nanomechanical resonators are fabricated from a high-stress $Si_3N_4$ layer, 90 nm thick and under an in-plane isotropic pre-stress of 1.06 GPa, deposited via low-pressure chemical vapor deposition (LPCVD) on a silicon substrate. The resonator's pattern was defined using electron beam (e-beam) lithography on a layer of e-beam resist (ARP6200-13), which was spin-coated onto the $Si_3N_4$ film. The pattern was then transferred into the $Si_3N_4$ thin film using reactive ion etching (RIE) with $CHF_3$ plasma. Afterwards, the resist was removed with hot dimethylformamide solution in a supersonic bath, followed by Piranha and diluted hydrofluoric acid cleaning to remove organic residues and surface oxides. Finally, the $Si_3N_4$ layer was released from the silicon substrate using cryogenic inductively couple plasma (ICP) etching with $SF_6$ to etch isotropically the silicon substrate, producing a 5 µm-wide undercut around each of our resonators [1].

All nanomechanical resonators studied in this work are made of $Si_3N_4$ deposited on the same wafer, which guarantees almost identical mechanical properties, with an initial isotropic stress $\sigma_0 = 1.06$ GPa, Young's modulus $E = 271$ GPa, Poisson's ratio $\nu = 0.23$, mass density $\rho = 3100$ kg/m$^3$. Furthermore, we estimate the intrinsic mechanical $Q$-factor of the nanoresoantors to be $Q_0 = [28000^{-1} + (6 \times 10^{10} h)^{-1}]^{-1} = 4527$ for $h = 90$ nm [2].

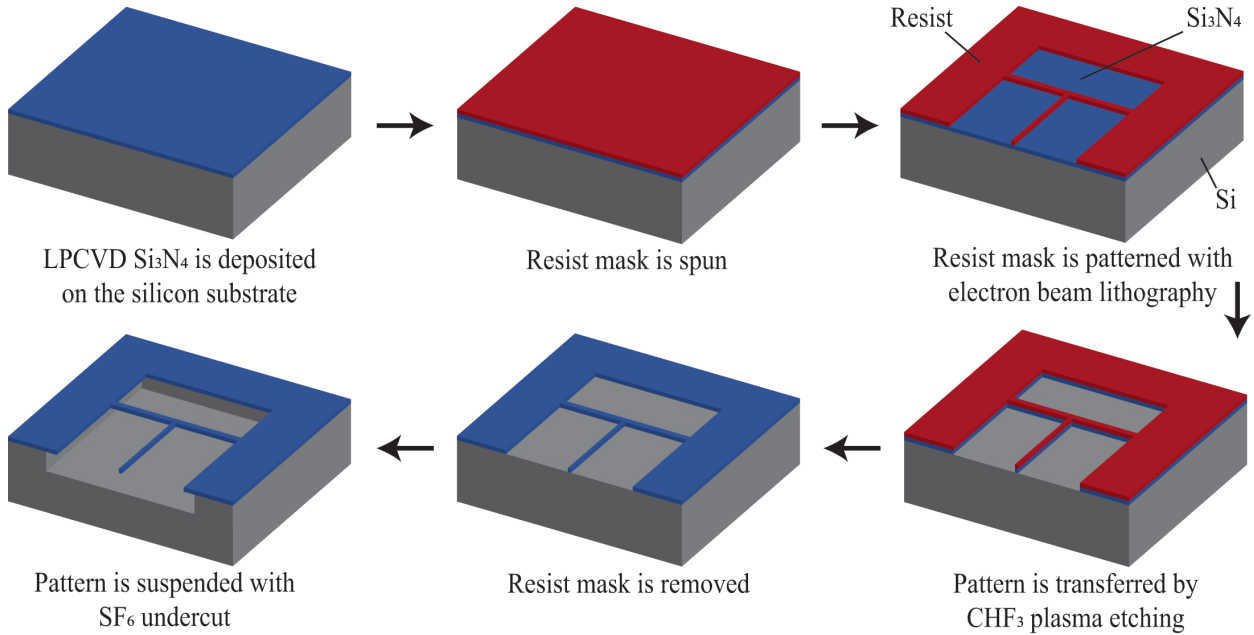

Figure S1: **Fabrication process of suspended $Si_3N_4$ nanomechanical resonators.**

## S2. Simultaneous measurement of different resonance modes

In order to detect the low-order out-of-plane modes simultaneously under intermodal coupling, we focus the laser beam of a Polytec MSA400 laser Doppler vibrometer (LDV) on the central string at $1/12L$ away from the support, as shown in Figure S2, away from the nodes of the vibrational modes. We use a Zurich lock-in amplifier to demodulate the measured signal with the driven frequency $f$ and its higher harmonics $nf$ ($n=2, 3, ...$). Next, we convert the measured displacements of different modes to their corresponding maximum displacements $q_1, q_2, ..., q_n$ according to different mode shapes. This is done to facilitate comparison with analytical and finite element (FE)-based reduced order models (ROMs). Moreover, since intermodal couplings normally occur at relatively large amplitudes that might exceed the limited measurement range of the LDV, it is also beneficial to perform measurements at locations with smaller amplitudes, away from the peaks of various modes.

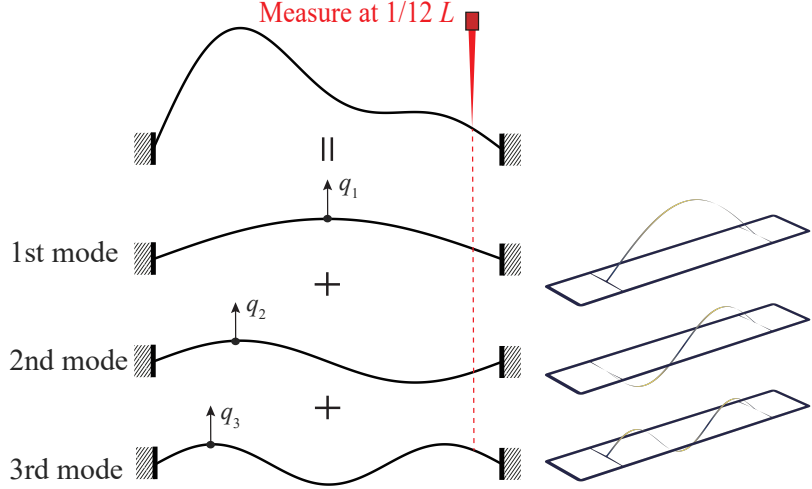

Figure S2: **Simultaneous measurement of multiple modes by LDV.** $q_1$, $q_2$ and $q_3$ are the maximum displacements of the three lowest out-of-plane modes, respectively. The mode shapes obtained by eigenfrequency analysis of the FE model are brought next to the schematic ones.

## S3. Two-mode dispersive coupling

Our devices can be treated as simply supported string resonators, so the effective mass of the low-order out-of-plane modes are approximately the same [5]. Since our resonators are flat, we only consider cubic nonlinear terms in our mass-normalized equations of motion [4]:

$$\begin{aligned}
\ddot{q}_1 + c_1\dot{q}_1 + \omega_1^2 q_1 + k_{111}^{(1)}q_1^3 + k_{112}^{(1)}q_1^2 q_2 + k_{122}^{(1)}q_1 q_2^2 + k_{222}^{(1)}q_2^3 &= F_{\text{exc}}\sin(2\pi ft) \\
\ddot{q}_2 + c_2\dot{q}_2 + \omega_2^2 q_2 + k_{111}^{(2)}q_1^3 + k_{112}^{(2)}q_1^2 q_2 + k_{122}^{(2)}q_1 q_2^2 + k_{222}^{(2)}q_2^3 &= F'_{\text{exc}}\sin(4\pi ft).
\end{aligned} \quad \text{(S1)}$$

in which $q_1$ and $q_2$ represent the displacement of the first and second mode, $c_1$ and $c_2$ represent the mass-normalized damping coefficients, $\omega_1$ and $\omega_2$ are the eigenfrequencies, $k$ with superscripts and subscripts are mass-normalized nonlinear stiffness. $F_{\text{exc}}\sin(2\pi ft)$ is the effective harmonic drive with excitation frequency $f$. Due to the non-uniformity at the interface between the piezo shaker and our sample, the second mode is also under a weak but non-negligible harmonic excitation, modeled as $F'_{\text{exc}}\sin(4\pi ft)$ in the simulations. To qualitatively account for this effect, $F'_{\text{exc}}$ is set to $10^{-6}\cdot F_{\text{exc}}$ during numerical simulations. It is worth noting that in Eq. (S1), not all terms are resonant under the interaction between two modes of a string with a resonance frequency ratio close to two. To recover the resonant terms, we assume harmonic motions of the form $q_1 = A_1\cos(\omega t)$ and $q_2 = A_2\cos(2\omega t)$ as a first approximation, and simplify Eq. (S1) as:

$$\begin{aligned}
\ddot{q}_1 + c_1\dot{q}_1 + \omega_1^2 q_1 + k_{111}^{(1)}q_1^3 + k_{122}^{(1)}q_1 q_2^2 &= F_{\text{exc}}\sin(2\pi ft) \\
\ddot{q}_2 + c_2\dot{q}_2 + \omega_2^2 q_2 + k_{112}^{(2)}q_1^2 q_2 + k_{222}^{(2)}q_2^3 &= F'_{\text{exc}}\sin(4\pi ft),
\end{aligned} \quad \text{(S2)}$$

where the resonant dispersive coupling term could be written as $\gamma = k_{122}^{(1)} = k_{112}^{(2)}$, and the mass-normalized Duffing constant of the first and second modes are $\beta_1 = k_{111}^{(1)}$ and $\beta_2 = k_{222}^{(2)}$, respectively. For simplicity, we rewrite Eq. (S2) as:

$$\ddot{q}_1 + c_1\dot{q}_1 + \omega_1^2 q_1 + \beta_1 q_1^3 + \gamma q_1 q_2^2 = F_{\text{exc}} \sin(2\pi f t)$$
$$\ddot{q}_2 + c_2\dot{q}_2 + \omega_2^2 q_2 + \gamma q_1^2 q_2 + \beta_2 q_2^3 = F'_{\text{exc}} \sin(4\pi f t). \tag{S3}$$

To understand the influence of the dispersive coupling on the kink observed in the frequency response curves of the main text, next we use the harmonic balance method (HBM), and assume the solution to be of the form: $q_1 = A_1 \cos(\omega t)$ and $q_2 = A_2 \cos(2\omega t)$. Replacing these harmonic responses in Eq. (S3), gives:

$$\omega^2 = \omega_1^2 + \frac{3}{4}\beta_1 A_1^2 + \frac{1}{2}\gamma A_2^2 \tag{S4a}$$

$$\omega^2 = \frac{1}{4}\omega_2^2 + \frac{1}{8}\gamma A_1^2 + \frac{3}{16}\beta_2 A_2^2. \tag{S4b}$$

We assume the second mode is activated at $\omega_{1,c}$, where $A_2 = 0$ and $A_1 = A_{1,c}$. Then Eqs. (S4a), (S4b) become:

$$\omega_{1,c}^2 = \omega_1^2 + \frac{3}{4}\beta_1 A_{1,c}^2 \tag{S5a}$$

$$\omega_{1,c}^2 = \frac{1}{4}\omega_2^2 + \frac{1}{8}\gamma A_{1,c}^2. \tag{S5b}$$

There are two unknown parameters ($\omega_{1,c}$ and $A_{1,c}$) in Eqs. (S5a), (S5b) and by solving them, we can analytically derive the position where the first mode's backbone undergoes a kink, as shown in Figure S3a:

$$\omega_{1,c} = \sqrt{\frac{\omega_2^2 - \frac{2\gamma}{3\beta_1}\omega_1^2}{4 - \frac{2\gamma}{3\beta_1}}} \tag{S6a}$$

$$A_{1,c} = \sqrt{\frac{\omega_2^2 - 4\omega_1^2}{3\beta_1 - \frac{1}{2}\gamma}}. \tag{S6b}$$

Furthermore, by eliminating $A_2$ from Eqs. (S4a), (S4b) we can obtain the backbone of the first mode during mode coupling as follows:

$$\omega^2 = \frac{\omega_1^2 - \frac{2\gamma}{3\beta_2}\omega_2^2}{1 - \frac{8\gamma}{3\beta_2}} + \frac{\frac{3}{4}\beta_1 - \frac{\gamma^2}{3\beta_2}}{1 - \frac{8\gamma}{3\beta_2}} A_1^2. \tag{S7}$$

We define the coefficient of $A_1^2$ term as effective Duffing constant $\beta_{1,\text{eff}}$ of the first mode due to two-mode dispersive coupling.

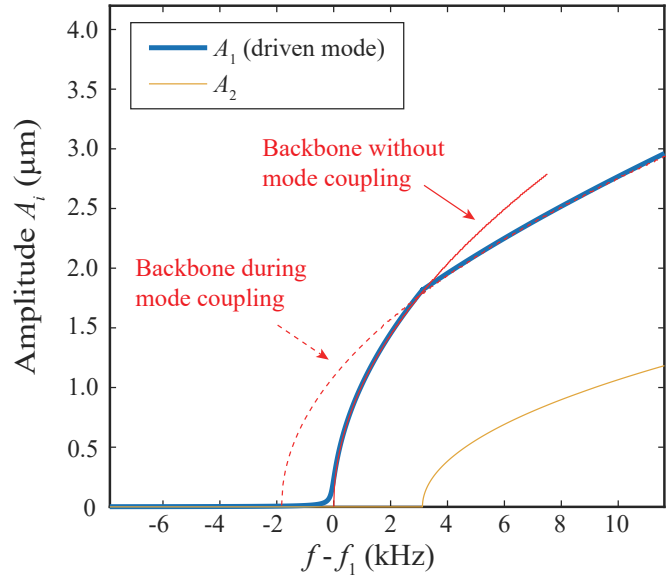

Figure S3: **Derivation of the effective backbone curve under two-mode dispersive coupling.** The simulated response curve of a forward frequency sweep near the first mode's resonance, that activates the second mode due to dispersive coupling.

3

## S4. Fitting nonlinear coefficients by a single frequency sweep

We developed a fitting strategy for the fast characterization of nonlinear properties of high-$Q$ nanomechanical resonators that undergo dispersive coupling. In Figure S4a, we present the measured frequency response curve for a device with $w_s = 1\,\mu m$, $L_s = 90\,\mu m$, and $\theta = 0$, where we observed the activation of the second mode by driving the first mode in the nonlinear regime. It is worth noticing that in high-$Q$ Duffing resonators, frequency sweeps in the nonlinear regime can bring the oscillations to their high-amplitude stable branches, which are in close proximity to their backbones [6]. Consequently, we can approximately fit for unknown coefficients in the backbone expressions using the frequency responses. We select the data marked as red circles in Figure S4a to fit the surface described by Eq. (S4a), as shown in Figure S4b. With the measured $\omega^2$, $A_1^2$ and $A_2^2$ from one frequency sweep, we can simultaneously fit for the three unknowns $\omega_1$, $\beta_1$ and $\gamma$. The values of $\omega_2$ and $\beta_2$ can be obtained by fitting the frequency responses obtained by driving the system around $\omega_2$ in the nonlinear regime.

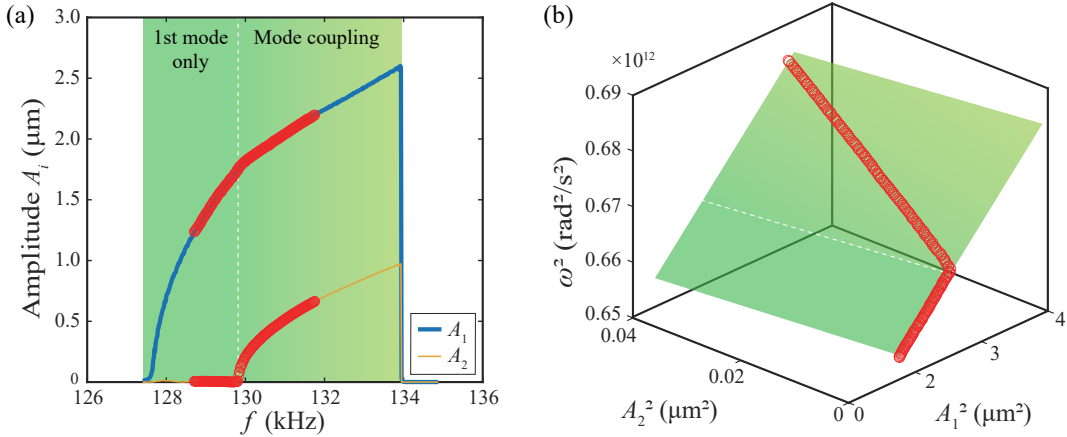

Figure S4: **Fitting the nonlinear coefficients of a 2-DOF nonlinear dynamical system that exhibits dispersive coupling in a single frequency sweep.** (a) The measured coupled response between the lowest two modes are from the device with $w_s = 1\,\mu m$, $L_s = 90\,\mu m$, and $\theta = 0$. The bold blue line represents the frequency response of the driven mode, which is demodulated with the frequency ($f$). The yellow line represents the frequency response of the second mode, which is demodulated with twice the driven frequency ($2f$). The selected data for fitting are marked as red circles. (b) The fitted surface is described by Eq. (S4a) with the selected data.

## S5. Dispersive coupling between the lowest two modes of a softly clamped string

By using Lagrange equations, we obtain the equations of motion of a string resonator with a pair of in-plane springs $k_{in}$ at both of its ends [4], as shown in Figure S4.

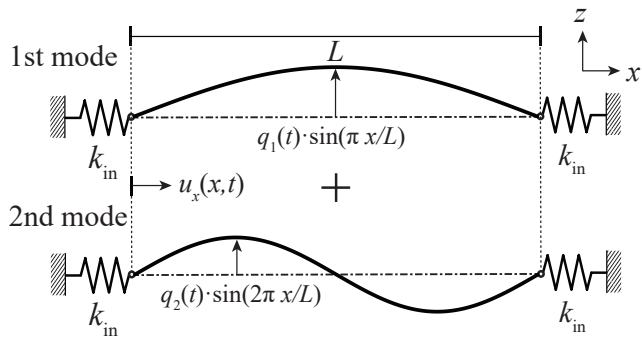

Figure S4: **Simplified model for a string resonator with soft-clamping supports.** We model the influence from soft-clamping supports as a pair of in-plane springs $k_{in}$ at both ends of the central string. The mode shapes are approximated as sinusoidal functions, which are the close form solutions for simply supported strings.

We assume mode shapes along the $x$-axis and coordinates as the function of time $t$. The out-of-plane and in-plane displacement for a string between the two modes as shown in Figure S4 can be written as:

4

$$u_z(x,t) = q_1(t)\sin\left(\frac{\pi}{L}x\right) + q_2(t)\sin\left(2\frac{\pi}{L}x\right), \tag{S8a}$$

$$u_x(x,t) = u_0\left(\frac{2x}{L} - 1\right) + \sum_{i=1}^{N} u_i(t)\sin\left(\frac{i\pi x}{L}\right) + u_\mathrm{a}(t)\left(1 - \frac{2x}{L}\right), \tag{S8b}$$

where $u$ and $w$ are the displacements of the string in $x$ (in-plane) and $z$ (out-of-plane) directions, respectively. $u_i(x,t)$ represents the in-plane counterpart of a stressed simply supported string during the out-of-plane motion $u_z(x,t)$. $u_\mathrm{a}(t)$ is the additional in-plane motion caused by the finite in-plane stiffness $k_\mathrm{in}$, and $u_0$ represents the initial stress in $x$ direction of the string:

$$u_0 = \frac{(1-\nu)\sigma_0 L}{2E}, \tag{S9}$$

$u_\mathrm{a}(t)$ in Eq. (S8b) is the additional in-plane motion that is considered to ensure satisfaction of the boundary conditions, which is in contrast to the ones in simply supported strings ($k_\mathrm{in} \to \infty$, $u_\mathrm{a}(t) \to 0$). $u_\mathrm{a}(t)$ can be obtained as:

$$\begin{aligned}k_\mathrm{in} u_\mathrm{a}(t) &= \frac{EA}{L}\int_0^L \varepsilon \, \mathrm{d}x \\ &= \frac{EA}{L}\int_0^L \left[\frac{\partial u_x}{\partial x} + \frac{1}{2}\left(\frac{\partial u_z}{\partial x}\right)^2\right]\mathrm{d}x,\end{aligned} \tag{S10}$$

where $A = hw$ is the area of the string's cross section and $\varepsilon$ is the strain along the $x$ direction. Next, by substituting Eqs. (S8a), (S8b) in Eq. (S10), one obtains:

$$u_\mathrm{a}(t) = \frac{\pi^2 \left[q_1^2(t) + 4q_2^2(t)\right]}{8L}\left(1 + \frac{k_\mathrm{in}L}{2EA}\right)^{-1}. \tag{S11}$$

The strain energy of a string can then be written as:

$$U_\mathrm{s} = \frac{1}{2}\int_0^L EA\varepsilon^2 \mathrm{d}x. \tag{S12}$$

Furthermore, the energy stored in the two in-plane springs $k_\mathrm{in}$ is:

$$U_\mathrm{k} = 2 \times \frac{1}{2}k_\mathrm{in} u_\mathrm{a}^2(t). \tag{S13}$$

Additionally, the kinetic energy of the string neglecting the in-plane inertia is given by:

$$T = \frac{1}{2}\rho A \int_0^L \left(\frac{\partial u_z}{\partial t}\right)^2 \mathrm{d}x. \tag{S14}$$

Using Eqs. (S12) - (S14), the Lagrange equations can be constructed as:

$$\frac{\mathrm{d}}{\mathrm{d}t}\left(\frac{\partial T}{\partial \dot{\boldsymbol{q}}}\right) - \frac{\partial T}{\partial \boldsymbol{q}} + \frac{\partial U}{\partial \boldsymbol{q}} = 0, \tag{S15}$$

where $\boldsymbol{q} = [q_1(t), q_2(t), u_1(t), u_2(t), ..., u_i(t)], i = 3, 4, ..., N$, is the vector that includes the generalized coordinates. Since the in-plane inertia has been neglected, after substituting the potential energy $U = U_\mathrm{s} + U_\mathrm{k}$ and the kinetic energy $T$ in Eq. (S14), we have a system of nonlinear equations consisting of two differential equations associated with the generalized coordinates $q_1(t)$ and $q_2(t)$, and $N$ algebraic equations in terms of $u_i(t)$. By solving the $N$ algebraic equations we can determine $u_i(t)$ in terms of $q_1(t)$ and $q_2(t)$:

$$\begin{aligned}u_1(t) &= -\frac{\pi q_1(t)q_2(t)}{L}\\ u_2(t) &= -\frac{\pi q_1^2(t)}{8L}\\ u_3(t) &= -\frac{\pi q_1(t)q_2(t)}{3L}\\ u_4(t) &= -\frac{\pi q_2^2(t)}{4L}\\ u_5(t) &= u_6(t) = ... = u_N(t) = 0,\end{aligned} \tag{S16}$$

As such we reduce the $N + 2$ nonlinear equations to two coupled equations as follows:

$$m_{\text{eff},1}\ddot{q}_1 + c_1\dot{q}_1 + k_{1,1}q_1 + k_{3,1}q_1^3 + k_{c,1}q_1 q_2^2 = 0 \tag{S17a}$$

$$m_{\text{eff},2}\ddot{q}_2 + c_2\dot{q}_2 + k_{1,2}q_2 + k_{3,2}q_2^3 + k_{c,2}q_1^2 q_2 = 0. \tag{S17b}$$

where $m_{\text{eff},1} = m_{\text{eff},2} = \rho A L/2$, and

$$\omega_1^2 = \frac{k_{1,1}}{m_{\text{eff},1}} = \frac{\pi^2(1-\nu)\sigma_0}{\rho L^2}\left(1+\frac{2EA}{k_{\text{in}}L}\right)^{-1}, \tag{S18a}$$

$$\omega_2^2 = \frac{k_{1,2}}{m_{\text{eff},2}} = \frac{4\pi^2(1-\nu)\sigma_0}{\rho L^2}\left(1+\frac{2EA}{k_{\text{in}}L}\right)^{-1}, \tag{S18b}$$

Accordingly, we can obtain the resonance frequency of the lowest two mechanical modes as:

$$f_1 = \frac{1}{2L}\sqrt{\frac{1}{\rho}\frac{(1-\nu)\sigma_0}{1+\frac{2EA}{k_{\text{in}}L}}} = \frac{1}{2L}\sqrt{\frac{\sigma_b}{\rho}}, \tag{S19a}$$

$$f_2 = \frac{1}{L}\sqrt{\frac{1}{\rho}\frac{(1-\nu)\sigma_0}{1+\frac{2EA}{k_{\text{in}}L}}} = \frac{1}{L}\sqrt{\frac{\sigma_b}{\rho}}, \tag{S19b}$$

where $\sigma_b$ is the stress in the string in $x$ direction after the release etch [3]:

$$\sigma_b = (1-\nu)\sigma_0\left(1+\frac{2EA}{k_{\text{in}}L}\right)^{-1}. \tag{S20}$$

Finally, the analytical derivation of the mass-normalized nonlinear coefficients $\beta_1, \beta_2, \gamma$, which are used to plot the solid lines of Figure 3b of the main text, are:

$$\beta_1 = \frac{k_{3,1}}{m_{\text{eff},1}} = \frac{\pi^4 E}{4\rho L^4}\left(1+\frac{2EA}{k_{\text{in}}L}\right)^{-1}, \tag{S21a}$$

$$\beta_2 = \frac{k_{3,2}}{m_{\text{eff},2}} = \frac{4\pi^4 E}{\rho L^4}\left(1+\frac{2EA}{k_{\text{in}}L}\right)^{-1}, \tag{S21b}$$

$$\gamma = \frac{k_{c,1}}{m_{\text{eff},1}} = \frac{k_{c,2}}{m_{\text{eff},2}} = \frac{\pi^4 E}{\rho L^4}\left(1+\frac{2EA}{k_{\text{in}}L}\right)^{-1}, \tag{S21c}$$

It is worth noting that $(1+2EA/k_{\text{in}}L)^{-1}$ serves as a tuning factor introduced by the finite in-plane stiffness $k_{\text{in}}$, which rescales $\sigma_b$, $\omega_1^2$, $\omega_2^2$, $\beta_1$, $\beta_2$, and the mass-normalized coupling strength $\gamma$ of a string with pre-tension $(1-\nu)\sigma_0$.

## S6. Parameters obtained by finite element-based reduced-order modeling

The parameters used for the simulated frequency responses presented in the main text are provided here and were obtained from finite element (FE)-based reduced-order models (ROMs) [4]. Table S1 presents the mass-normalized parameters that describe the dispersive coupling between the first two modes of devices with different values of $L_s$. These parameters can be directly used to construct Eq. (S2) for simulating the response curves shown in Fig. 2b and Fig. 3c of the main text. For the simulations in Fig. 2b, the effective harmonic drive level is set to $F_{\text{exc}} = 65.65$ (m/s$^2$). For Fig. 3c, a device-specific drive level is used, given by $4.00 \times 10^{-4} \cdot 4\pi^2 f_1^2 h$ (m/s$^2$) for devices with different $L_s$. $f_1$ denotes the eigenfrequency of the first mode, and $h$ represents the thickness (90 nm) of our devices. The non-resonant terms are omitted, as they have a negligible effect on the systems' dynamical response [7]. It is worth mentioning that for a potential of the form $U = \frac{1}{2}\gamma q_1^2(t)q_2^2(t)$, $k_{122}^{(1)} = k_{112}^{(2)} = \gamma$. Yet the slight difference observed between $k_{122}^{(1)}$ and $k_{112}^{(2)}$ in Table S1 comes from FE-based ROM construction. We use $k_{122}^{(1)}$ as the value for $\gamma$ from FE-based ROMs presented in Fig. 3b.

The parameters used for simulating the response curves shown in Fig. 4c and d are shown in Table S2. Apart from eigenfrequencies $f_i$ and quality factors $Q_i$, all coefficients for cubic nonlinear terms ($\sum_{l=1}^{5}\sum_{m=l}^{5}\sum_{n=m}^{5} k_{lmn}^{(i)} q_l q_m q_n$) are presented including the non-resonant terms. The effective harmonic drive here is set to be $F_{\text{exc}} = 474.56$ (m/s$^2$).

Table S1: Mass-normalized parameters for dispersive coupling between the first two modes of devices with varying $L_\text{s}$

| $L_\text{s}$(μm) | $f_1$(Hz) | $f_2$(Hz) | $Q_1$ | $Q_2$ | $k^{(1)}_{111}$(m$^{-2}$s$^{-2}$) | $k^{(2)}_{222}$(m$^{-2}$s$^{-2}$) | $k^{(1)}_{122}$(m$^{-2}$s$^{-2}$) | $k^{(2)}_{112}$(m$^{-2}$s$^{-2}$) |
|---|---|---|---|---|---|---|---|---|
| 130 | $1.04 \times 10^5$ | $2.11 \times 10^5$ | $2.06 \times 10^5$ | $1.00 \times 10^5$ | $8.16 \times 10^{21}$ | $1.29 \times 10^{23}$ | $3.53 \times 10^{22}$ | $3.46 \times 10^{22}$ |
| 120 | $1.10 \times 10^5$ | $2.22 \times 10^5$ | $2.18 \times 10^5$ | $1.08 \times 10^5$ | $9.05 \times 10^{21}$ | $1.43 \times 10^{23}$ | $3.91 \times 10^{22}$ | $3.82 \times 10^{22}$ |
| 110 | $1.16 \times 10^5$ | $2.34 \times 10^5$ | $2.31 \times 10^5$ | $1.18 \times 10^5$ | $1.02 \times 10^{22}$ | $1.60 \times 10^{23}$ | $4.37 \times 10^{22}$ | $4.26 \times 10^{22}$ |
| 100 | $1.24 \times 10^5$ | $2.50 \times 10^5$ | $2.47 \times 10^5$ | $1.29 \times 10^5$ | $1.16 \times 10^{22}$ | $1.81 \times 10^{23}$ | $4.95 \times 10^{22}$ | $4.82 \times 10^{22}$ |
| 90 | $1.33 \times 10^5$ | $2.68 \times 10^5$ | $2.66 \times 10^5$ | $1.44 \times 10^5$ | $1.34 \times 10^{22}$ | $2.10 \times 10^{23}$ | $5.78 \times 10^{22}$ | $5.61 \times 10^{22}$ |
| 80 | $1.45 \times 10^5$ | $2.92 \times 10^5$ | $2.90 \times 10^5$ | $1.62 \times 10^5$ | $1.60 \times 10^{22}$ | $2.50 \times 10^{23}$ | $6.91 \times 10^{22}$ | $6.68 \times 10^{22}$ |
| 70 | $1.60 \times 10^5$ | $3.22 \times 10^5$ | $3.22 \times 10^5$ | $1.87 \times 10^5$ | $1.98 \times 10^{22}$ | $3.07 \times 10^{23}$ | $8.57 \times 10^{22}$ | $8.24 \times 10^{22}$ |
| 60 | $1.81 \times 10^5$ | $3.63 \times 10^5$ | $3.65 \times 10^5$ | $2.23 \times 10^5$ | $2.56 \times 10^{22}$ | $3.95 \times 10^{23}$ | $1.12 \times 10^{23}$ | $1.07 \times 10^{23}$ |
| 50 | $2.11 \times 10^5$ | $4.24 \times 10^5$ | $4.29 \times 10^5$ | $2.77 \times 10^5$ | $3.56 \times 10^{22}$ | $5.44 \times 10^{23}$ | $1.58 \times 10^{23}$ | $1.49 \times 10^{23}$ |
| 40 | $2.60 \times 10^5$ | $5.22 \times 10^5$ | $5.30 \times 10^5$ | $3.66 \times 10^5$ | $5.53 \times 10^{22}$ | $8.35 \times 10^{23}$ | $2.49 \times 10^{23}$ | $2.33 \times 10^{23}$ |
| 30 | $3.50 \times 10^5$ | $7.01 \times 10^5$ | $7.03 \times 10^5$ | $5.29 \times 10^5$ | $1.02 \times 10^{23}$ | $1.53 \times 10^{24}$ | $4.65 \times 10^{23}$ | $4.32 \times 10^{23}$ |

Table S2: Mass-normalized parameters for five-mode coupling of the device with $L_\text{s} = 50$μm

| Mode number $(i)$ | 1 | 2 | 3 | 4 | 5 |
|---|---|---|---|---|---|
| $f_i$ (Hz) | $2.11 \times 10^5$ | $4.24 \times 10^5$ | $6.39 \times 10^5$ | $8.58 \times 10^5$ | $1.08 \times 10^6$ |
| $Q_i$ | $4.28 \times 10^5$ | $2.76 \times 10^5$ | $1.75 \times 10^5$ | $1.16 \times 10^5$ | $8.22 \times 10^4$ |
| $k^{(i)}_{111}$ (m$^{-2}$s$^{-2}$) | $3.56 \times 10^{22}$ | $2.08 \times 10^{18}$ | $-1.66 \times 10^{21}$ | $4.85 \times 10^{18}$ | $-3.35 \times 10^{21}$ |
| $k^{(i)}_{112}$ (m$^{-2}$s$^{-2}$) | $6.64 \times 10^{18}$ | $1.49 \times 10^{23}$ | $1.19 \times 10^{19}$ | $-1.88 \times 10^{21}$ | $1.78 \times 10^{19}$ |
| $k^{(i)}_{113}$ (m$^{-2}$s$^{-2}$) | $-5.12 \times 10^{21}$ | $1.15 \times 10^{19}$ | $3.29 \times 10^{23}$ | $2.17 \times 10^{19}$ | $-3.25 \times 10^{21}$ |
| $k^{(i)}_{114}$ (m$^{-2}$s$^{-2}$) | $1.36 \times 10^{19}$ | $-1.91 \times 10^{21}$ | $2.16 \times 10^{19}$ | $5.81 \times 10^{23}$ | $3.19 \times 10^{19}$ |
| $k^{(i)}_{115}$ (m$^{-2}$s$^{-2}$) | $-1.08 \times 10^{22}$ | $2.10 \times 10^{19}$ | $-3.23 \times 10^{21}$ | $3.19 \times 10^{19}$ | $9.05 \times 10^{23}$ |
| $k^{(i)}_{122}$ (m$^{-2}$s$^{-2}$) | $1.58 \times 10^{23}$ | $2.57 \times 10^{19}$ | $-2.64 \times 10^{21}$ | $3.46 \times 10^{19}$ | $-5.85 \times 10^{21}$ |
| $k^{(i)}_{123}$ (m$^{-2}$s$^{-2}$) | $5.92 \times 10^{21}$ | $-2.73 \times 10^{22}$ | $-4.98 \times 10^{22}$ | $2.51 \times 10^{22}$ | $6.64 \times 10^{20}$ |
| $k^{(i)}_{124}$ (m$^{-2}$s$^{-2}$) | $-8.68 \times 10^{21}$ | $2.15 \times 10^{22}$ | $2.62 \times 10^{22}$ | $-8.59 \times 10^{22}$ | $4.39 \times 10^{22}$ |
| $k^{(i)}_{125}$ (m$^{-2}$s$^{-2}$) | $-6.38 \times 10^{21}$ | $1.10 \times 10^{22}$ | $-3.03 \times 10^{20}$ | $4.38 \times 10^{22}$ | $-1.34 \times 10^{23}$ |
| $k^{(i)}_{133}$ (m$^{-2}$s$^{-2}$) | $3.51 \times 10^{23}$ | $3.50 \times 10^{19}$ | $-5.54 \times 10^{22}$ | $7.19 \times 10^{19}$ | $-1.23 \times 10^{22}$ |
| $k^{(i)}_{134}$ (m$^{-2}$s$^{-2}$) | $-7.02 \times 10^{21}$ | $2.38 \times 10^{22}$ | $4.84 \times 10^{22}$ | $-1.4 \times 10^{23}$ | $3.16 \times 10^{21}$ |
| $k^{(i)}_{135}$ (m$^{-2}$s$^{-2}$) | $-1.05 \times 10^{22}$ | $9.62 \times 10^{19}$ | $2.24 \times 10^{22}$ | $2.05 \times 10^{20}$ | $-2.1 \times 10^{23}$ |
| $k^{(i)}_{144}$ (m$^{-2}$s$^{-2}$) | $6.22 \times 10^{23}$ | $6.41 \times 10^{19}$ | $-2.73 \times 10^{22}$ | $2.28 \times 10^{20}$ | $-5.09 \times 10^{22}$ |
| $k^{(i)}_{145}$ (m$^{-2}$s$^{-2}$) | $3.74 \times 10^{21}$ | $4.17 \times 10^{22}$ | $2.74 \times 10^{21}$ | $-2.39 \times 10^{22}$ | $1.24 \times 10^{23}$ |
| $k^{(i)}_{155}$ (m$^{-2}$s$^{-2}$) | $9.70 \times 10^{23}$ | $1.05 \times 10^{20}$ | $-4.04 \times 10^{22}$ | $2.24 \times 10^{20}$ | $-2.78 \times 10^{23}$ |
| $k^{(i)}_{222}$ (m$^{-2}$s$^{-2}$) | $1.05 \times 10^{19}$ | $5.44 \times 10^{23}$ | $3.76 \times 10^{19}$ | $-1.96 \times 10^{22}$ | $4.75 \times 10^{19}$ |
| $k^{(i)}_{223}$ (m$^{-2}$s$^{-2}$) | $-3.56 \times 10^{21}$ | $6.8 \times 10^{19}$ | $1.25 \times 10^{24}$ | $7.71 \times 10^{19}$ | $-3.17 \times 10^{22}$ |
| $k^{(i)}_{224}$ (m$^{-2}$s$^{-2}$) | $3.23 \times 10^{19}$ | $-6.13 \times 10^{22}$ | $7.79 \times 10^{19}$ | $2.21 \times 10^{24}$ | $1.31 \times 10^{20}$ |
| $k^{(i)}_{225}$ (m$^{-2}$s$^{-2}$) | $-6.13 \times 10^{21}$ | $1.17 \times 10^{20}$ | $-3.13 \times 10^{22}$ | $1.3 \times 10^{20}$ | $3.44 \times 10^{24}$ |
| $k^{(i)}_{233}$ (m$^{-2}$s$^{-2}$) | $3.38 \times 10^{19}$ | $1.26 \times 10^{24}$ | $1.50 \times 10^{20}$ | $-2.21 \times 10^{22}$ | $1.59 \times 10^{20}$ |
| $k^{(i)}_{234}$ (m$^{-2}$s$^{-2}$) | $1.99 \times 10^{22}$ | $-1.65 \times 10^{22}$ | $-8.02 \times 10^{22}$ | $6.72 \times 10^{22}$ | $1.00 \times 10^{23}$ |
| $k^{(i)}_{235}$ (m$^{-2}$s$^{-2}$) | $9.25 \times 10^{20}$ | $-8.31 \times 10^{22}$ | $-4.23 \times 10^{22}$ | $9.99 \times 10^{22}$ | $1.10 \times 10^{23}$ |
| $k^{(i)}_{244}$ (m$^{-2}$s$^{-2}$) | $6.49 \times 10^{19}$ | $2.24 \times 10^{24}$ | $1.66 \times 10^{20}$ | $-2.46 \times 10^{23}$ | $2.79 \times 10^{20}$ |
| $k^{(i)}_{245}$ (m$^{-2}$s$^{-2}$) | $3.27 \times 10^{22}$ | $1.97 \times 10^{22}$ | $9.85 \times 10^{22}$ | $-6.64 \times 10^{22}$ | $-3.13 \times 10^{23}$ |
| $k^{(i)}_{255}$ (m$^{-2}$s$^{-2}$) | $9.83 \times 10^{19}$ | $3.49 \times 10^{24}$ | $2.80 \times 10^{20}$ | $-1.04 \times 10^{23}$ | $7.20 \times 10^{20}$ |
| $k^{(i)}_{333}$ (m$^{-2}$s$^{-2}$) | $-1.6 \times 10^{22}$ | $1.14 \times 10^{20}$ | $2.73 \times 10^{24}$ | $2.55 \times 10^{20}$ | $-8.13 \times 10^{22}$ |
| $k^{(i)}_{334}$ (m$^{-2}$s$^{-2}$) | $6.78 \times 10^{19}$ | $-2.24 \times 10^{22}$ | $2.94 \times 10^{20}$ | $4.93 \times 10^{24}$ | $3.09 \times 10^{20}$ |
| $k^{(i)}_{335}$ (m$^{-2}$s$^{-2}$) | $-1.29 \times 10^{22}$ | $1.57 \times 10^{20}$ | $-2.56 \times 10^{23}$ | $2.98 \times 10^{20}$ | $7.68 \times 10^{24}$ |
| $k^{(i)}_{344}$ (m$^{-2}$s$^{-2}$) | $-2.41 \times 10^{22}$ | $1.75 \times 10^{20}$ | $4.94 \times 10^{24}$ | $5.21 \times 10^{20}$ | $-7.66 \times 10^{22}$ |
| $k^{(i)}_{345}$ (m$^{-2}$s$^{-2}$) | $-2.98 \times 10^{20}$ | $9.41 \times 10^{22}$ | $3.35 \times 10^{22}$ | $-2.1 \times 10^{23}$ | $-9.32 \times 10^{22}$ |
| $k^{(i)}_{355}$ (m$^{-2}$s$^{-2}$) | $-3.54 \times 10^{22}$ | $2.83 \times 10^{20}$ | $7.72 \times 10^{24}$ | $5.37 \times 10^{20}$ | $-7.11 \times 10^{23}$ |
| $k^{(i)}_{444}$ (m$^{-2}$s$^{-2}$) | $2.19 \times 10^{20}$ | $-7.36 \times 10^{22}$ | $5.59 \times 10^{20}$ | $8.61 \times 10^{24}$ | $1.28 \times 10^{21}$ |
| $k^{(i)}_{445}$ (m$^{-2}$s$^{-2}$) | $-4.03 \times 10^{22}$ | $2.97 \times 10^{20}$ | $-7.65 \times 10^{22}$ | $9.3 \times 10^{20}$ | $1.36 \times 10^{25}$ |
| $k^{(i)}_{455}$ (m$^{-2}$s$^{-2}$) | $2.02 \times 10^{20}$ | $-1.00 \times 10^{23}$ | $5.29 \times 10^{20}$ | $1.36 \times 10^{25}$ | $1.51 \times 10^{21}$ |
| $k^{(i)}_{555}$ (m$^{-2}$s$^{-2}$) | $-6.96 \times 10^{22}$ | $1.23 \times 10^{21}$ | $-2.07 \times 10^{23}$ | $2.53 \times 10^{21}$ | $2.10 \times 10^{25}$ |

## S7. Measurement data of the coupled dynamics for the lowest two modes

In Table S3, we provide all the measured values (solid diamonds and triangles) in Fig. 3b, d and e of the main text. The resonance frequencies $f_1$, $f_2$ and nonlinear coefficients $\beta_1$, $\beta_2$, $\gamma$ are fitted from measured frequency response curves using the method introduced in Supplementary Information S4. The onset frequency $f_{1,c}$ and the onset amplitude $A_{1,c}$ of the coupled response are directly determined from the measured frequency response curves.

Table S3: Measurement data of Fig. 3b, c, d and e

| $L_s$(µm) | $f_1$(Hz) | $f_2$(Hz) | $\beta_1$(m$^{-2}$s$^{-2}$) | $\beta_2$(m$^{-2}$s$^{-2}$) | $\gamma$(m$^{-2}$s$^{-2}$) | $f_{1,c}$(Hz) | $A_{1,c}$(µm) |
|---|---|---|---|---|---|---|---|
| 130 | 93700 | 189958 | $5.21 \times 10^{21}$ | $3.76 \times 10^{22}$ | $1.80 \times 10^{22}$ | 97157 | 2.48 |
| 110 | 107676 | 217257 | $6.81 \times 10^{21}$ | $5.93 \times 10^{22}$ | $2.40 \times 10^{22}$ | 110179 | 2.30 |
| 90 | 127701 | 256128 | $8.60 \times 10^{21}$ | $1.49 \times 10^{23}$ | $3.80 \times 10^{22}$ | 129810 | 1.77 |
| 70 | 156532 | 314299 | $1.17 \times 10^{22}$ | $1.83 \times 10^{23}$ | $3.44 \times 10^{22}$ | 158005 | 1.44 |
| 50 | 207892 | 416427 | $2.01 \times 10^{22}$ | $2.92 \times 10^{23}$ | $6.07 \times 10^{22}$ | 208676 | 0.92 |
| 30 | 347320 | 694724 | $7.73 \times 10^{22}$ | $1.66 \times 10^{24}$ | $9.33 \times 10^{22}$ | 347739 | 0.46 |

## S8. Multi-mode dispersive coupling

To study the influence of cascaded dispersive couplings in a string resonator, here we derive the expression of the effective Duffing constant of the first mode undergoing couplings with higher-order modes. We only consider coupling terms that are derived from the interaction potential $U = \frac{1}{2}\gamma_{1,i}q_1^2(t)q_i^2(t)$ ($i \geq 2$). This approximation neglects the interactions of the coupled mode with modes other than the first mode. Similar to our approach in deriving Eqs. (S4a) and (S4b), if we consider the effective backbone of the first mode when coupled to the $i$th mode, we have:

$$4\omega^2 = 4\omega_{1,\text{eff}}^{(i-1)^2} + 3\beta_{1,\text{eff}}^{(i-1)}A_1^2 + 2\gamma_{1,i}A_i^2, \tag{S22a}$$

$$4i^2\omega^2 = 4\omega_i^2 + 2\gamma_{1,i}A_1^2 + 3\beta_i A_i^2, \tag{S22b}$$

where $\beta_i$ is the Duffing constant of the $i$th mode, and $\gamma_{1,i}$ is the dispersive coupling strength between the first and $i$th mode. By eliminating $A_i$ from Eqs. (S22a), (S22b), we can derive the new effective Duffing constant of the first mode:

$$\beta_{1,\text{eff}}^{(i)} = \frac{9\beta_{1,\text{eff}}^{(i-1)}\beta_i - 4\gamma_{1,i}^2}{12\beta_i - 8i^2\gamma_{1,i}}. \tag{S23}$$

By separating $\frac{3}{4}\beta_{1,\text{eff}}^{(i-1)}$ from Eq. (S23), we can quantitatively visualize the influence of successive coupled modes on the Duffing constant:

$$\beta_{1,\text{eff}}^{(i)} = \frac{3}{4}\beta_{1,\text{eff}}^{(i-1)} + \frac{3i^2\beta_{1,\text{eff}}^{(i-1)}\gamma_{1,i} - 2\gamma_{1,i}^2}{6\beta_i - 4i^2\gamma_{1,i}}. \tag{S24}$$

In Table S4, we present the dynamical parameters obtained by the FE-based ROMs for the calculation of Eq. (S24). These parameters are selected from Table S2, where $\beta_i$ corresponds to $k_{iii}^{(i)}$, and $\gamma_{1,i}$ to $k_{1ii}^{(1)}$. The effective Duffing constants $\beta_{1,\text{eff}}^{(i)}$ during cascaded dispersive couplings are calculated using Eq. (S24), based on the Duffing constants of different modes and their dispersive coupling coefficients with the first mode. The increasing values of $(\beta_{1,\text{eff}}^{(i)} - \beta_1)/\beta_1$ demonstrate that the effective Duffing constant of the first mode can be significantly tuned via successive couplings to other vibrational modes.

Table S4: Tuning of the effective Duffing constant $\beta_{1,\text{eff}}^{(i)}$ by successive dispersive mode coupling

| Mode number ($i$) | 1 | 2 | 3 | 4 | 5 |
|---|---|---|---|---|---|
| $\beta_i$(m$^{-2}$s$^{-2}$) | $3.76 \times 10^{22}$ | $5.44 \times 10^{23}$ | $2.73 \times 10^{24}$ | $8.61 \times 10^{24}$ | $2.10 \times 10^{25}$ |
| $\gamma_{1,i}$(m$^{-2}$s$^{-2}$) | — | $1.58 \times 10^{23}$ | $3.53 \times 10^{23}$ | $6.22 \times 10^{23}$ | $9.70 \times 10^{23}$ |
| $\beta_{1,\text{eff}}^{(i)}$(m$^{-2}$s$^{-2}$) | — | $5.72 \times 10^{22}$ | $1.24 \times 10^{23}$ | $3.39 \times 10^{23}$ | $1.04 \times 10^{24}$ |

To more quantitatively show the tunability of the driven mode's response by incorporating multiple coupled modes, in Figure S5, we simulate the frequency response of our string resonators with $w_s = 1\,\mu\text{m}$, $\theta = 0$ and varying $L_s$ with FE-based ROMs up to five modes (see Table S5, S2 and S6 for the simulation parameters). We can see the flattening of the frequency response of the first mode of the device with $L_s = 50\,\mu\text{m}$ (same with Fig. 4d in the main text) and

even the drop in the one with $L_s = 30\,\mu m$, which represent the effective Duffing constant $\beta_{1,\text{eff}} \to \infty$ and $\beta_{1,\text{eff}} < 0$, respectively. The additional tuning effect compared to the values shown in Table S4 is attributed to the inclusion of all coupling terms related to cubic geometric nonlinearity in the FE-based ROMs, as opposed to the assumption for Eq. (S24). Both analytical and numerical investigations suggest that, through geometric design and multi-mode interaction, one can engineer the coupled dynamical response of a resonator to a large extent.

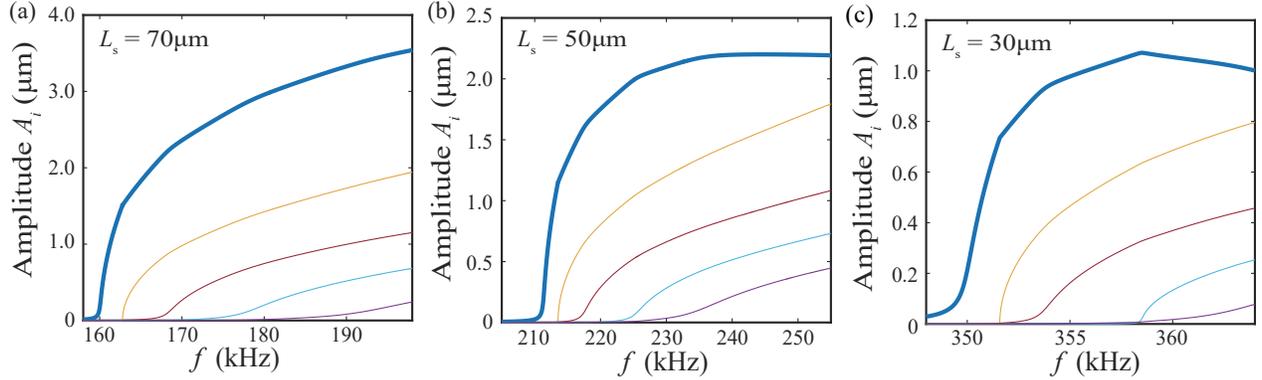

Figure S5: **Simulated frequency response curves based on FE-based ROMs under multi-mode coupling of devices with different support lengths.** The first modes of all three devices is driven by forward frequency sweeps at the same drive level. The blue, yellow, ochre, cyan and purple lines represent the first to the fifth modes, respectively.

# Supplementary References

[1] M. Xu, D. Shin, P. M. Sberna, R. van der Kolk, A. Cupertino, M. A. Bessa, and R. A. Norte, High-strength amorphous silicon carbide for nanomechanics, Advanced Materials 36, 2306513 (2024).

[2] L. G. Villanueva and S. Schmid, Evidence of surface loss as ubiquitous limiting damping mechanism in sin micro- and nanomechanical resonators, Physical Review Letters 113, 227201 (2014).

[3] Z. Li, M. Xu, R. A. Norte, A. M. Aragón, F. van Keulen, F. Alijani, and P. G. Steeneken, Tuning the $Q$-factor of nanomechanical string resonators by torsion support design, Applied Physics Letters 122, 013501 (2023).

[4] Z. Li, M. Xu, R. A. Norte, A. M. Aragón, P. G. Steeneken, and F. Alijani, Strain engineering of nonlinear nanoresonators from hardening to softening. Communications Physics, 7, 53 (2024).

[5] B. Hauer, C. Doolin, K. Beach, J. Davis, A general procedure for thermomechanical calibration of nano/micro-mechanical resonators. Annals of Physics, 339, 181–207 (2013).

[6] A. H. Nayfeh and D. T. Mook, Nonlinear oscillations (John Wiley & Sons, 2008).

[7] Y. Yang, E. Ng, P. Polunin, Y. Chen, S. Strachan, V. Hong, C. H. Ahn, O. Shoshani, S. Shaw, M. Dykman, and T. Kenny, Experimental investigation on mode coupling of bulk mode silicon mems resonators, in 2015 28th IEEE International Conference on Micro Electro Mechanical Systems (MEMS) (IEEE, 2015) pp. 1008–1011.

Table S5: Mass-normalized parameters for five-mode coupling of the device with $L_\text{s} = 70\text{µm}$

| Mode number ($i$) | 1 | 2 | 3 | 4 | 5 |
|---|---|---|---|---|---|
| $f_i$ (Hz) | $1.60 \times 10^5$ | $3.22 \times 10^5$ | $4.87 \times 10^5$ | $6.57 \times 10^5$ | $8.33 \times 10^5$ |
| $Q_i$ | $3.22 \times 10^5$ | $1.87 \times 10^5$ | $1.12 \times 10^5$ | $7.26 \times 10^4$ | $5.08 \times 10^4$ |
| $k_{111}^{(i)}$ (m$^{-2}$s$^{-2}$) | $1.98 \times 10^{22}$ | $4.67 \times 10^{18}$ | $-8.76 \times 10^{20}$ | $8.83 \times 10^{18}$ | $-1.77 \times 10^{21}$ |
| $k_{112}^{(i)}$ (m$^{-2}$s$^{-2}$) | $1.36 \times 10^{19}$ | $8.24 \times 10^{22}$ | $4.16 \times 10^{19}$ | $-1.68 \times 10^{20}$ | $6.82 \times 10^{19}$ |
| $k_{113}^{(i)}$ (m$^{-2}$s$^{-2}$) | $-2.63 \times 10^{21}$ | $4.09 \times 10^{19}$ | $1.83 \times 10^{23}$ | $8.13 \times 10^{19}$ | $-1.27 \times 10^{20}$ |
| $k_{114}^{(i)}$ (m$^{-2}$s$^{-2}$) | $2.68 \times 10^{19}$ | $-2.00 \times 10^{20}$ | $8.01 \times 10^{19}$ | $3.23 \times 10^{23}$ | $1.34 \times 10^{20}$ |
| $k_{115}^{(i)}$ (m$^{-2}$s$^{-2}$) | $-5.65 \times 10^{21}$ | $6.73 \times 10^{19}$ | $-1.10 \times 10^{20}$ | $1.33 \times 10^{20}$ | $5.04 \times 10^{23}$ |
| $k_{122}^{(i)}$ (m$^{-2}$s$^{-2}$) | $8.57 \times 10^{22}$ | $5.21 \times 10^{19}$ | $-9.50 \times 10^{19}$ | $1.13 \times 10^{20}$ | $-8.60 \times 10^{20}$ |
| $k_{123}^{(i)}$ (m$^{-2}$s$^{-2}$) | $2.06 \times 10^{21}$ | $-8.46 \times 10^{21}$ | $-1.89 \times 10^{22}$ | $2.03 \times 10^{22}$ | $5.39 \times 10^{20}$ |
| $k_{124}^{(i)}$ (m$^{-2}$s$^{-2}$) | $3.10 \times 10^{21}$ | $-9.02 \times 10^{21}$ | $2.12 \times 10^{22}$ | $-3.40 \times 10^{22}$ | $3.54 \times 10^{22}$ |
| $k_{125}^{(i)}$ (m$^{-2}$s$^{-2}$) | $-2.58 \times 10^{21}$ | $7.26 \times 10^{21}$ | $-1.54 \times 10^{20}$ | $3.55 \times 10^{22}$ | $-5.18 \times 10^{22}$ |
| $k_{133}^{(i)}$ (m$^{-2}$s$^{-2}$) | $1.91 \times 10^{23}$ | $1.26 \times 10^{20}$ | $-2.87 \times 10^{22}$ | $2.53 \times 10^{20}$ | $-1.34 \times 10^{21}$ |
| $k_{134}^{(i)}$ (m$^{-2}$s$^{-2}$) | $2.80 \times 10^{21}$ | $1.96 \times 10^{22}$ | $-1.80 \times 10^{22}$ | $-5.27 \times 10^{22}$ | $-9.52 \times 10^{20}$ |
| $k_{135}^{(i)}$ (m$^{-2}$s$^{-2}$) | $-1.16 \times 10^{21}$ | $4.26 \times 10^{20}$ | $1.48 \times 10^{22}$ | $8.60 \times 10^{20}$ | $-7.71 \times 10^{22}$ |
| $k_{144}^{(i)}$ (m$^{-2}$s$^{-2}$) | $3.39 \times 10^{23}$ | $2.22 \times 10^{20}$ | $-9.81 \times 10^{21}$ | $4.38 \times 10^{20}$ | $-1.90 \times 10^{22}$ |
| $k_{145}^{(i)}$ (m$^{-2}$s$^{-2}$) | $-9.77 \times 10^{20}$ | $3.45 \times 10^{22}$ | $-8.67 \times 10^{20}$ | $-8.26 \times 10^{21}$ | $-4.59 \times 10^{22}$ |
| $k_{155}^{(i)}$ (m$^{-2}$s$^{-2}$) | $5.30 \times 10^{23}$ | $3.44 \times 10^{20}$ | $-1.38 \times 10^{22}$ | $6.98 \times 10^{20}$ | $-1.44 \times 10^{23}$ |
| $k_{222}^{(i)}$ (m$^{-2}$s$^{-2}$) | $1.83 \times 10^{19}$ | $3.07 \times 10^{23}$ | $6.89 \times 10^{19}$ | $-1.06 \times 10^{22}$ | $9.94 \times 10^{19}$ |
| $k_{223}^{(i)}$ (m$^{-2}$s$^{-2}$) | $-7.76 \times 10^{20}$ | $1.68 \times 10^{20}$ | $7.05 \times 10^{23}$ | $3.30 \times 10^{20}$ | $-1.16 \times 10^{22}$ |
| $k_{224}^{(i)}$ (m$^{-2}$s$^{-2}$) | $1.08 \times 10^{20}$ | $-3.28 \times 10^{22}$ | $3.28 \times 10^{20}$ | $1.25 \times 10^{24}$ | $5.31 \times 10^{20}$ |
| $k_{225}^{(i)}$ (m$^{-2}$s$^{-2}$) | $-1.32 \times 10^{21}$ | $2.56 \times 10^{20}$ | $-1.15 \times 10^{22}$ | $5.25 \times 10^{20}$ | $1.94 \times 10^{24}$ |
| $k_{233}^{(i)}$ (m$^{-2}$s$^{-2}$) | $1.18 \times 10^{20}$ | $7.10 \times 10^{23}$ | $3.73 \times 10^{20}$ | $-4.67 \times 10^{21}$ | $5.78 \times 10^{20}$ |
| $k_{234}^{(i)}$ (m$^{-2}$s$^{-2}$) | $1.78 \times 10^{22}$ | $8.10 \times 10^{21}$ | $6.81 \times 10^{21}$ | $2.84 \times 10^{22}$ | $8.92 \times 10^{22}$ |
| $k_{235}^{(i)}$ (m$^{-2}$s$^{-2}$) | $9.31 \times 10^{20}$ | $-3.08 \times 10^{22}$ | $-1.53 \times 10^{22}$ | $8.87 \times 10^{22}$ | $4.51 \times 10^{22}$ |
| $k_{244}^{(i)}$ (m$^{-2}$s$^{-2}$) | $2.15 \times 10^{20}$ | $1.26 \times 10^{24}$ | $6.49 \times 10^{20}$ | $-1.31 \times 10^{23}$ | $1.06 \times 10^{21}$ |
| $k_{245}^{(i)}$ (m$^{-2}$s$^{-2}$) | $2.95 \times 10^{22}$ | $-3.99 \times 10^{21}$ | $8.88 \times 10^{22}$ | $-2.13 \times 10^{22}$ | $-3.69 \times 10^{22}$ |
| $k_{255}^{(i)}$ (m$^{-2}$s$^{-2}$) | $3.31 \times 10^{20}$ | $1.97 \times 10^{24}$ | $1.03 \times 10^{21}$ | $-3.76 \times 10^{22}$ | $1.57 \times 10^{21}$ |
| $k_{333}^{(i)}$ (m$^{-2}$s$^{-2}$) | $-8.64 \times 10^{21}$ | $1.69 \times 10^{20}$ | $1.54 \times 10^{24}$ | $3.57 \times 10^{20}$ | $-4.35 \times 10^{22}$ |
| $k_{334}^{(i)}$ (m$^{-2}$s$^{-2}$) | $2.41 \times 10^{20}$ | $-4.98 \times 10^{21}$ | $6.81 \times 10^{20}$ | $2.79 \times 10^{24}$ | $1.18 \times 10^{21}$ |
| $k_{335}^{(i)}$ (m$^{-2}$s$^{-2}$) | $-2.31 \times 10^{21}$ | $5.84 \times 10^{20}$ | $-1.37 \times 10^{23}$ | $1.16 \times 10^{21}$ | $4.35 \times 10^{24}$ |
| $k_{344}^{(i)}$ (m$^{-2}$s$^{-2}$) | $-8.78 \times 10^{21}$ | $6.28 \times 10^{20}$ | $2.80 \times 10^{24}$ | $1.22 \times 10^{21}$ | $-1.66 \times 10^{22}$ |
| $k_{345}^{(i)}$ (m$^{-2}$s$^{-2}$) | $1.13 \times 10^{21}$ | $8.57 \times 10^{22}$ | $-9.04 \times 10^{21}$ | $-5.30 \times 10^{22}$ | $3.96 \times 10^{22}$ |
| $k_{355}^{(i)}$ (m$^{-2}$s$^{-2}$) | $-1.22 \times 10^{22}$ | $9.94 \times 10^{20}$ | $4.38 \times 10^{24}$ | $1.96 \times 10^{21}$ | $-3.79 \times 10^{23}$ |
| $k_{444}^{(i)}$ (m$^{-2}$s$^{-2}$) | $2.66 \times 10^{20}$ | $-3.95 \times 10^{22}$ | $7.21 \times 10^{20}$ | $4.87 \times 10^{24}$ | $1.38 \times 10^{21}$ |
| $k_{445}^{(i)}$ (m$^{-2}$s$^{-2}$) | $-1.45 \times 10^{22}$ | $1.06 \times 10^{21}$ | $-1.65 \times 10^{22}$ | $2.03 \times 10^{21}$ | $7.72 \times 10^{24}$ |
| $k_{455}^{(i)}$ (m$^{-2}$s$^{-2}$) | $6.62 \times 10^{20}$ | $-3.58 \times 10^{22}$ | $1.97 \times 10^{21}$ | $7.74 \times 10^{24}$ | $3.25 \times 10^{21}$ |
| $k_{555}^{(i)}$ (m$^{-2}$s$^{-2}$) | $-3.71 \times 10^{22}$ | $1.37 \times 10^{21}$ | $-1.10 \times 10^{23}$ | $2.35 \times 10^{21}$ | $1.19 \times 10^{25}$ |

Table S6: Mass-normalized parameters for five-mode coupling of the device with $L_s = 30$μm

| Mode number ($i$) | 1 | 2 | 3 | 4 | 5 |
|---|---|---|---|---|---|
| $f_i$ (Hz) | $3.50 \times 10^5$ | $7.01 \times 10^5$ | $1.05 \times 10^6$ | $1.41 \times 10^6$ | $1.77 \times 10^6$ |
| $Q_i$ | $7.03 \times 10^5$ | $5.29 \times 10^5$ | $3.75 \times 10^5$ | $2.67 \times 10^4$ | $1.96 \times 10^5$ |
| $k_{111}^{(i)}$ (m$^{-2}$s$^{-2}$) | $1.02 \times 10^{23}$ | $-2.74 \times 10^{18}$ | $-4.95 \times 10^{21}$ | $-2.96 \times 10^{18}$ | $-9.93 \times 10^{21}$ |
| $k_{112}^{(i)}$ (m$^{-2}$s$^{-2}$) | $-6.26 \times 10^{18}$ | $4.32 \times 10^{23}$ | $-3.35 \times 10^{19}$ | $-8.46 \times 10^{21}$ | $-6.20 \times 10^{19}$ |
| $k_{113}^{(i)}$ (m$^{-2}$s$^{-2}$) | $-1.56 \times 10^{22}$ | $-3.40 \times 10^{19}$ | $9.47 \times 10^{23}$ | $-6.99 \times 10^{19}$ | $-1.52 \times 10^{22}$ |
| $k_{114}^{(i)}$ (m$^{-2}$s$^{-2}$) | $-9.79 \times 10^{18}$ | $-8.45 \times 10^{21}$ | $-7.21 \times 10^{19}$ | $1.67 \times 10^{24}$ | $-1.18 \times 10^{20}$ |
| $k_{115}^{(i)}$ (m$^{-2}$s$^{-2}$) | $-3.25 \times 10^{22}$ | $-5.88 \times 10^{19}$ | $-1.51 \times 10^{22}$ | $-1.18 \times 10^{20}$ | $2.59 \times 10^{24}$ |
| $k_{122}^{(i)}$ (m$^{-2}$s$^{-2}$) | $4.66 \times 10^{23}$ | $-2.73 \times 10^{19}$ | $-1.25 \times 10^{22}$ | $-9.21 \times 10^{19}$ | $-2.54 \times 10^{22}$ |
| $k_{123}^{(i)}$ (m$^{-2}$s$^{-2}$) | $2.42 \times 10^{22}$ | $-1.12 \times 10^{23}$ | $-1.95 \times 10^{23}$ | $5.14 \times 10^{22}$ | $2.30 \times 10^{21}$ |
| $k_{124}^{(i)}$ (m$^{-2}$s$^{-2}$) | $-3.84 \times 10^{22}$ | $8.72 \times 10^{22}$ | $5.41 \times 10^{22}$ | $-3.38 \times 10^{23}$ | $9.05 \times 10^{22}$ |
| $k_{125}^{(i)}$ (m$^{-2}$s$^{-2}$) | $-2.37 \times 10^{22}$ | $3.77 \times 10^{22}$ | $2.29 \times 10^{20}$ | $8.85 \times 10^{22}$ | $-5.21 \times 10^{23}$ |
| $k_{133}^{(i)}$ (m$^{-2}$s$^{-2}$) | $1.03 \times 10^{24}$ | $-1.07 \times 10^{20}$ | $-1.67 \times 10^{23}$ | $-1.99 \times 10^{20}$ | $-5.50 \times 10^{22}$ |
| $k_{134}^{(i)}$ (m$^{-2}$s$^{-2}$) | $-2.80 \times 10^{22}$ | $4.73 \times 10^{22}$ | $1.90 \times 10^{23}$ | $-5.27 \times 10^{23}$ | $7.06 \times 10^{21}$ |
| $k_{135}^{(i)}$ (m$^{-2}$s$^{-2}$) | $-4.94 \times 10^{22}$ | $-3.60 \times 10^{20}$ | $8.05 \times 10^{22}$ | $-7.11 \times 10^{20}$ | $-8.01 \times 10^{23}$ |
| $k_{144}^{(i)}$ (m$^{-2}$s$^{-2}$) | $1.82 \times 10^{24}$ | $-1.94 \times 10^{20}$ | $-9.66 \times 10^{22}$ | $-1.44 \times 10^{20}$ | $-1.78 \times 10^{23}$ |
| $k_{145}^{(i)}$ (m$^{-2}$s$^{-2}$) | $1.53 \times 10^{22}$ | $8.20 \times 10^{22}$ | $4.73 \times 10^{21}$ | $-4.32 \times 10^{22}$ | $4.88 \times 10^{23}$ |
| $k_{155}^{(i)}$ (m$^{-2}$s$^{-2}$) | $2.84 \times 10^{24}$ | $-2.99 \times 10^{20}$ | $-1.45 \times 10^{23}$ | $-5.64 \times 10^{20}$ | $-8.42 \times 10^{23}$ |
| $k_{222}^{(i)}$ (m$^{-2}$s$^{-2}$) | $-6.66 \times 10^{17}$ | $1.53 \times 10^{24}$ | $1.58 \times 10^{19}$ | $-5.73 \times 10^{22}$ | $-2.52 \times 10^{19}$ |
| $k_{223}^{(i)}$ (m$^{-2}$s$^{-2}$) | $-1.42 \times 10^{22}$ | $-6.79 \times 10^{19}$ | $3.53 \times 10^{24}$ | $-2.73 \times 10^{20}$ | $-1.10 \times 10^{23}$ |
| $k_{224}^{(i)}$ (m$^{-2}$s$^{-2}$) | $-9.55 \times 10^{19}$ | $-1.79 \times 10^{23}$ | $-2.78 \times 10^{20}$ | $6.22 \times 10^{24}$ | $-4.53 \times 10^{20}$ |
| $k_{225}^{(i)}$ (m$^{-2}$s$^{-2}$) | $-2.45 \times 10^{22}$ | $-1.43 \times 10^{20}$ | $-1.09 \times 10^{23}$ | $-4.69 \times 10^{20}$ | $9.66 \times 10^{24}$ |
| $k_{233}^{(i)}$ (m$^{-2}$s$^{-2}$) | $-1.07 \times 10^{20}$ | $3.57 \times 10^{24}$ | $-1.54 \times 10^{20}$ | $-8.89 \times 10^{22}$ | $-5.57 \times 10^{20}$ |
| $k_{234}^{(i)}$ (m$^{-2}$s$^{-2}$) | $3.56 \times 10^{22}$ | $-7.10 \times 10^{22}$ | $-3.26 \times 10^{23}$ | $2.69 \times 10^{23}$ | $1.79 \times 10^{23}$ |
| $k_{235}^{(i)}$ (m$^{-2}$s$^{-2}$) | $1.83 \times 10^{21}$ | $-2.99 \times 10^{23}$ | $-1.72 \times 10^{23}$ | $1.77 \times 10^{23}$ | $4.34 \times 10^{23}$ |
| $k_{244}^{(i)}$ (m$^{-2}$s$^{-2}$) | $-1.90 \times 10^{20}$ | $6.32 \times 10^{24}$ | $-5.64 \times 10^{20}$ | $-7.21 \times 10^{23}$ | $-9.98 \times 10^{20}$ |
| $k_{245}^{(i)}$ (m$^{-2}$s$^{-2}$) | $5.74 \times 10^{22}$ | $7.79 \times 10^{22}$ | $1.73 \times 10^{23}$ | $-2.73 \times 10^{23}$ | $-1.14 \times 10^{24}$ |
| $k_{255}^{(i)}$ (m$^{-2}$s$^{-2}$) | $-2.99 \times 10^{20}$ | $9.84 \times 10^{24}$ | $-8.65 \times 10^{20}$ | $-3.63 \times 10^{23}$ | $-9.52 \times 10^{20}$ |
| $k_{333}^{(i)}$ (m$^{-2}$s$^{-2}$) | $-4.63 \times 10^{22}$ | $1.56 \times 10^{20}$ | $7.66 \times 10^{24}$ | $4.08 \times 10^{20}$ | $-2.38 \times 10^{23}$ |
| $k_{334}^{(i)}$ (m$^{-2}$s$^{-2}$) | $-2.11 \times 10^{20}$ | $-8.88 \times 10^{22}$ | $-3.53 \times 10^{20}$ | $1.38 \times 10^{25}$ | $-1.08 \times 10^{21}$ |
| $k_{335}^{(i)}$ (m$^{-2}$s$^{-2}$) | $-5.29 \times 10^{22}$ | $-5.31 \times 10^{20}$ | $-7.49 \times 10^{23}$ | $-1.08 \times 10^{21}$ | $2.15 \times 10^{25}$ |
| $k_{344}^{(i)}$ (m$^{-2}$s$^{-2}$) | $-8.35 \times 10^{22}$ | $-5.76 \times 10^{20}$ | $1.39 \times 10^{25}$ | $-6.40 \times 10^{20}$ | $-3.06 \times 10^{23}$ |
| $k_{345}^{(i)}$ (m$^{-2}$s$^{-2}$) | $-1.97 \times 10^{21}$ | $1.60 \times 10^{23}$ | $1.39 \times 10^{23}$ | $-8.51 \times 10^{23}$ | $-3.92 \times 10^{23}$ |
| $k_{355}^{(i)}$ (m$^{-2}$s$^{-2}$) | $-1.25 \times 10^{23}$ | $-8.80 \times 10^{20}$ | $2.16 \times 10^{25}$ | $-1.75 \times 10^{21}$ | $-2.08 \times 10^{24}$ |
| $k_{444}^{(i)}$ (m$^{-2}$s$^{-2}$) | $4.52 \times 10^{20}$ | $-2.15 \times 10^{23}$ | $1.24 \times 10^{21}$ | $2.41 \times 10^{25}$ | $2.73 \times 10^{21}$ |
| $k_{445}^{(i)}$ (m$^{-2}$s$^{-2}$) | $-1.40 \times 10^{23}$ | $-9.59 \times 10^{20}$ | $-3.06 \times 10^{23}$ | $-1.11 \times 10^{21}$ | $3.80 \times 10^{25}$ |
| $k_{455}^{(i)}$ (m$^{-2}$s$^{-2}$) | $-5.95 \times 10^{20}$ | $-3.48 \times 10^{23}$ | $-1.81 \times 10^{21}$ | $3.81 \times 10^{25}$ | $-1.62 \times 10^{21}$ |
| $k_{555}^{(i)}$ (m$^{-2}$s$^{-2}$) | $-2.05 \times 10^{23}$ | $3.30 \times 10^{21}$ | $-6.15 \times 10^{23}$ | $5.49 \times 10^{21}$ | $5.86 \times 10^{25}$ |



\end{document}